\documentclass[letterpaper]{article} 
\usepackage{aaai24}  
\usepackage{times}  
\usepackage{helvet}  
\usepackage{courier}  
\usepackage[hyphens]{url}  
\usepackage{graphicx} 
\usepackage{natbib}  
\usepackage{caption} 
\usepackage{subcaption}
\usepackage{algorithm}
\usepackage{algorithmic}
\usepackage{tabularx}
\usepackage{multirow}
\usepackage{amsmath}
\usepackage{amssymb}
\usepackage{mathtools}
\usepackage{amsthm}
\usepackage{cleveref}
\usepackage{placeins}
\newcommand\given[1][]{\:#1\vert\:}

\usepackage{newfloat}
\usepackage{listings}
\DeclareCaptionStyle{ruled}{labelfont=normalfont,labelsep=colon,strut=off} 
\floatstyle{ruled}
\newfloat{listing}{tb}{lst}{}
\floatname{listing}{Listing}

\newcommand\blfootnote[1]{%
\begingroup
\renewcommand\thefootnote{}\footnote{#1}%
\addtocounter{footnote}{-1}%
\endgroup
}

\title{A Bayesian Spatial Model to Correct Under-Reporting\\in Urban Crowdsourcing}
\author{ Gabriel Agostini, Emma Pierson\textdagger, Nikhil Garg\textdagger\blfootnote{\textdagger  Co-senior author}}
\affiliations{Cornell Tech, New York, United States\\
gsagostini@infosci.cornell.edu, emma.pierson@cornell.edu, ngarg@cornell.edu}

\begin{document}

\maketitle

\begin{abstract}
	Decision-makers often observe the occurrence of events through a reporting process. City governments, for example, rely on resident reports to find and then resolve urban infrastructural problems such as fallen street trees, flooded basements, or rat infestations. Without additional assumptions, there is no way to distinguish events that occur but are not reported from events that truly did not occur--a fundamental problem in settings with positive-unlabeled data. Because disparities in reporting rates correlate with resident demographics, addressing incidents only on the basis of reports leads to systematic neglect in neighborhoods that are less likely to report events. We show how to overcome this challenge by leveraging the fact that events are \textit{spatially correlated}. Our framework uses a Bayesian spatial latent variable model to infer event occurrence probabilities and applies it to storm-induced flooding reports in New York City, further pooling results across multiple storms. We show that a model accounting for under-reporting and spatial correlation predicts future reports more accurately than other models, and further induces a more equitable set of inspections: its allocations better reflect the population and provide equitable service to non-white, less traditionally educated, and lower-income residents. This finding reflects heterogeneous reporting behavior learned by the model: reporting rates are higher in Census tracts with higher populations, proportions of white residents, and proportions of owner-occupied households. Our work lays the groundwork for more equitable proactive government services, even with disparate reporting behavior.
\end{abstract}

\section{Introduction}
\label{sec:intro}
Urban crowdsourcing is key to identifying and resolving problems such as fallen street trees and flooded basements, in both emergency and daily contexts. For example, New York City's {311} system received over 3 million service requests in 2021 \cite{nyc_open_data_311_nodate}. However, reporting is \textit{heterogeneous} -- different neighborhoods, even when facing similar problems, report problems at different rates \cite{liu_quantifying_2023,kontokosta_equity_2017,minkoff_nyc_2016, obrien_uncharted_2017}, and under-reporting often correlates with socioeconomic factors such as race, ethnicity, and income. If agencies primarily address incidents that are reported, then under-reporting leads to downstream disparities. This mistargeting of resources, especially when it results in inequity, is a substantial concern and is a stated priority research area for the Federal Emergency Management Agency (FEMA) in the United States \cite{FEMAlearningagenda}.

While previous work has quantified the magnitude of reporting disparities, they do not estimate the probability that an event truly occurred at each location, which is essential for resource allocation. A challenge in estimation is that, without further assumptions, \textit{unreported} incidents cannot be distinguished from incidents which \textit{did not occur}. This is analogous to \textit{positive-unlabeled} (PU) machine learning \cite{liu_building_2003, shanmugam2021quantifying}, where datapoints are either labeled positive or unlabeled, and the latter group consists of both true positives and negatives. 

PU learning problems are unsolvable without further assumptions on the data generating process~\cite{bekker_learning_2020}. How do we make progress? Our insight is that many urban phenomena are \textit{spatially correlated}, and we can use this correlation to distinguish under-reporting from true lack of event occurrence. For example, in our empirical application, we use reporting data for \textit{flooding} after a storm. If an area does not report flooding but all of its spatial neighbors do, that area is likely to have experienced flooding (but not reported it); conversely, if no neighbors report flooding, that area is not likely to have experienced flooding. Spatial correlation departs from the standard PU learning setup, where the data points are assumed to be independent.

To encode spatial correlations, we build on top of a spatial Bayesian model developed in the ecology literature \cite{spezia_spatial_2018}. The model uses a latent indicator variable at each location to encode whether an event truly occurs there; latent variables at adjacent locations are spatially correlated according to an Ising model \cite{onsager_crystal_1944}. If an event does occur, the probability it generates an (observed) report varies as a function of location demographics. In semi-synthetic simulations, we show that the model infers where events have truly occurred more accurately than baseline models. For example, the model significantly outperforms a Gaussian Process (GP) in terms of AUC with an improvement of $0.14$. 

Using this model, we develop a novel framework to identify non-reported events in urban crowdsourcing and show it leads to more efficient and equitable resource allocation. We apply the framework to street flooding reports after storms in New York City obtained from public 311 data \cite{nyc_open_data_311_nodate}. NYC uses such report data to allocate post-storm resources, such as addressing wood debris, clogged catch basins, or building water leaks \cite{city_of_new_york_severe_2023}. Storms cause severe damage; Hurricane Ida in 2021 was responsible for the largest rainfall hour in the city's history, over 7 billions of dollars in damage to infrastructure and the transportation system, and at least 13 deaths, heavily skewed along socioeconomic lines \cite{newman_43_2021}. 

Our framework---which accounts for both heterogeneous under-reporting and spatial correlation---outperforms baseline approaches in terms of \textit{efficiency}: using report data immediately after the storm, it better predicts reports made days later. Such prediction can facilitate a more timely allocation of resources. Our framework also leads to more \emph{equitable} allocation of inspections: the allocations are more in line with population proportions, as opposed to other models that are likely to rate minority and lower-income neighborhoods as lower priority for inspections, due to under-reporting. Finally, we show how to pool model estimates \textit{across} storms to learn historical heterogeneous reporting patterns: we find that reporting rates are higher in Census tracts with higher median incomes, proportions of white residents, and proportions of owner-occupied housing.

Overall, our work (a) leverages spatial correlation in a Bayesian machine learning model to overcome the positive-unlabeled challenge in crowdsourcing; (b) develops a framework to validate and apply the model for more efficient and equitable emergency management response to a given storm event, and to pool reports across storms to identify under-reporting patterns; (c) applies the framework to real-world data, showing that it can substantially improve both the efficiency and equity of government responses to crowdsourcing. Our framework can improve responses in other contexts with spatial correlation, such as in public health and power outages. Together, we provide and validate a novel approach to resource \textit{allocation} in the presence of under-reporting.

We further provide an open-source Python implementation of our approach for application in other contexts\footnote{Available at github.com/gsagostini/networks\_underreporting/}. 

\section{Related Work}
\label{sec:related}
Our work relates to multiple threads of prior literature from PU Learning, Bayesian methods in ecology, flood prediction, and urban crowdsourcing. 

Our setting is one with \textbf{positive-unlabeled} data. Without further assumptions, even the proportion of true positive points---the \emph{prevalence}---is unidentifiable because a positive-class, unlabeled data point is indistinguishable from a negative-class point. Hence, PU learning methods \textit{must} make further assumptions \cite{bekker_learning_2020}; e.g., a common assumption is that each true positive point has the same uniform probability of being labeled positive \cite{elkan_learning_2008}. Even this strong assumption, which often does not hold in real-world settings where the labeling probability is non-uniform (e.g. when reporting is heterogeneous), is not sufficient. In contrast, our work overcomes the challenge by leveraging \textit{spatial correlation}.

Methodologically, our work builds on approaches from the \textbf{ecology} literature, which seeks to count animal populations in the presence of detection errors \cite{heikkinen_fully_1994, sicacha-parada_accounting_2021, della_rocca_new_2022,xu2023reflections}. \citet{santos-fernandez_correcting_2021} fit a Bayesian model to correct for misreporting errors in coral detection, leveraging spatial correlation and individuals that analyze coral in multiple locations. Most relevant is work by \citet{spezia_spatial_2018}. Their model assumes that the true probability of animal species presence is described by an Ising model, and observed presence is described by a reporting process. We build a framework to effectively use and validate this approach in urban crowdsourcing, showing that the model is predictive of future reports, can guide equitable resource allocation, and can be pooled across multiple events. 

This work's specific empirical application -- urban flood detection -- is complementary to the substantial machine learning work on \textbf{flood prediction}, using precipitation and seasonal climate information and data from satellites or sensors (see \citet{mosavi2018flood} for a comprehensive review). \citet{mauerman_high-quality_2022}, for example, use a Bayesian latent variable model to predict seasonal floods in Bangladesh through the reconstruction of historical satellite data. \citet{agonafir2022understanding} study infrastructure correlates of flooding using 311 reports in NYC. We believe that reporting data is a valuable complementary data source to sensors and satellites, and is especially temporally and spatially granular in urban environments; however, for both efficiency and equity, it is important to quantify and correct heterogeneous under-reporting. Future work could incorporate such outside sensor data into our model. Our approach is also applicable to reporting contexts beyond flooding. 

There is a large literature quantifying disparities in \textbf{urban crowdsourcing}; a consistent challenge is disambiguating between low reporting rates and low ground truth rates. \citet{liu_quantifying_2023} show that time-stamped, duplicate reports about the same event can be used to identify the reporting process; \citet{doi:10.1177/0081175015576601} send researchers to neighborhoods to document ground-truth conditions. We contribute a method that leverages spatial correlation and, unlike other methods, \textit{predicts} the probability an event has occurred in each location.

Finally, our work relates to a much broader literature on methods to quantify and compensate for the effects of missing and imperfect data in inequality-related contexts, including healthcare, policing, education, and government inspections \cite{coston2021characterizing,rambachan2021identifying,movva2023coarse,franchi2023detecting,end_to_end_auditing,guerdan2023counterfactual,zink2023race,cai2020fair,pierson2020assessing,liu2023library,sidhika2023selectivelabels,obermeyer2019dissecting,kleinberg2018human,zanger2023risk,jung2018omitted,garg2021standardized,lakkaraju2017selective,arnold2022measuring}. This broader literature considers many types of missingness besides the PU-missingness we study here, and many types of identification approaches besides the spatial correlations leveraged here.

\section{Model, Inference, and Framework}
\label{methods}
Our model captures three characteristics common to many urban crowdsourcing systems: (a) the city does not observe \textit{ground truth} data (where incidents actually occurred), only \textit{reports}; (b) there is \textit{under-reporting}, i.e., not all incidents are reported, and under-reporting may be \textit{heterogeneous} across demographic groups; (c) incidents are \textit{spatially correlated}.

Formally, consider a network $G$ with $N$ nodes and adjacency matrix $E$. Each node $i$ has two binary state variables. First, $A_i\in \{-1, +1\}$ denotes the latent, ground-truth state; second, $T_i\in\{0, 1\}$ denotes the observed, reported state. In the flood setting, $A_i = 1$ if a flood occurred in that node and $-1$ if not, while $T_i=1$ denotes that there was a report for flooding at the node. We observe reports $T_i$ and the network $G$, but not incidents $A_i$.  

Our specific approach follows that of \citet{spezia_spatial_2018}. \textbf{Ground truth states} $A_1, \dots A_N$ are generated according to an Ising model with two real-valued parameters, $\theta_0$ and $\theta_1$, controlling the event \textit{incidence rate} and spatial \textit{correlation} respectively. The probability distribution of the vector $\vec{A}\in \{\pm 1\}^N$ is:
\begin{align}\label{eq:A_dist}
	\Pr(\vec{A})
	&= \dfrac{\exp\left(\theta_0\sum_{i} A_i + \theta_1\sum_{i, j} A_iA_j \cdot E_{ij}\right)}{\mathcal{Z}(\theta_0, \theta_1)}
\end{align}
{where $\mathcal{Z}(\theta_0, \theta_1)$ is an intractable \emph{partition function} ensuring the distribution is normalized}. As proven by \citet{besag_spatial_1974}, the conditional distribution for a single node $A_i$ given all other nodes, is, with positive spatially correlation $\theta_1 > 0$:
\begin{align}
	\Pr(A_i=1&\given{A}_{k}\;\forall\, k\neq i)\nonumber \\&= \dfrac{1}{1 + \exp\left(-2\left(\theta_0 + \theta_1\sum_j A_j\cdot E_{ij}\right)\right)} \label{eq:aconditional}
\end{align}

A \textbf{report} at node $i$ only depends on the incident state at $i$ and a \textit{reporting rate} $\psi_i$, i.e.,
\begin{align}
	\Pr(T_i=1 \given A_i=1) = \psi_i. \label{eq:tgivenA}
\end{align}
As in PU learning, we assume that there are no false positive reports: $\Pr\left(T_i=1\mid A_i=-1\right) = 0$.

We fit and compare two models for reporting rates $\psi_i$:
\begin{itemize}
	\item With \textbf{homogeneous reporting}, $\psi_i = \alpha$ is assumed constant across nodes.
	\item With \textbf{heterogeneous reporting}, report rates $\psi_i$ are a function of demographic factors of node $i$. That is, given $M$ node-specific features $X_{i1}\dots X_{iM}$, 
	\begin{equation}\label{eq:psi}
		\psi_i = \text{logit}^{-1}\left(\alpha_0 + \sum_{\ell=1}^M\alpha_{\ell} X_{i\ell}\right),
	\end{equation}
	where the coefficients $\alpha_0, \dots \alpha_M$ are learned latent parameters shared across nodes. 
\end{itemize}

\noindent We discuss some of the modeling choices in \Cref{sec:discussion}.

\subsection{Inference procedure}
Given reporting data $\{T_i\}$ and a spatial network with known edges $\{E_{ij}\}$, we use a Gibbs sampling MCMC procedure for posterior inference: namely, at each iteration, we draw each latent value from its conditional distribution given the current values of all the other variables. All variables are initialized at random. Model priors are in \Cref{sec:SI-model}.

We provide code with our submission -- we note that we modify the procedure of \citet{spezia_spatial_2018}, to speed up inference, such as by jointly sampling some of the parameters within the outer Gibbs routine and implementing dynamic step size optimization. We believe that the public Python code release will enable other practitioners in crowdsourcing or ecology settings to apply such methods.  

\textbf{Sampling $\theta_0$ and $\theta_1$: } The conditional distribution of $\theta_0$ and $\theta_1$ given all other variables depends only on $\vec{A}$. We cannot directly compute it due to the presence of the partition function $\mathcal Z$ in \cref{eq:A_dist} \cite{murray_mcmc_nodate}. This normalization constant is intractable, as it must be evaluated for $2^N$ values of the ground-truth vector $\vec{A}$.

We use the Single-Variable Exchange Algorithm (SVEA) to circumvent this difficulty \cite{moller_efficient_2006}. The SVEA is a Metropolis-Hasting type sampling algorithm that introduces an auxiliary variable $\vec{w}$ to cancel two terms with the partition function when computing the acceptance ratio. To do so, $\vec{w}$ must be sampled from the same distribution family as $\vec{A}$. We generate auxiliary variables from the Ising model distribution in \cref{eq:A_dist} using the Swendsen-Wang algorithm with 50 burn-in samples \cite{swendsen_nonuniversal_1987, wolff_collective_1989}. This is an efficient method to sample from an Ising Model with $\theta_1>0$ \cite{park_rapid_2017,cooper_mixing_2000}.

\textbf{Sampling $A_i$:} We sample each of the $A_i$ through Gibbs sampling. The conditional distribution of $A_i$ depends on $\theta_0$, $\theta_1$, $T_i$, and $A_j$ for $j$ such that $E_{ij}=1$. If the corresponding $T_i=1$, then the no false positives assumption leads to $\Pr(A_i=1 \given \cdot) = 1$. Otherwise, $\Pr(A_i\given \cdot)$ is the conditional probability implied by \cref{eq:aconditional,eq:tgivenA}. 

\textbf{Sampling $\psi_i$: } In the \textit{homogeneous reporting model}, we fit a single parameter $\alpha$ to describe the reporting rate. Given a beta prior, the conditional distribution of $\alpha$ is a beta distribution depending on the numbers of incidents that 1) occurred and are reported and 2) occurred and are not reported.

In the \textit{heterogeneous reporting model}, the conditional distribution for the coefficients $\alpha_0, \dots \alpha_M$ can be found by fitting a Bayesian logistic regression of the reports $T_i$ on the demographic features, restricted to the nodes for which the current latent ground truth parameters are positive ($A_i = 1$). We compute $\psi_i$ following \cref{eq:psi}. 

\textbf{Sampling hyper-parameters.} We draw $3$ chains with 40,000 samples each, after 20,000 burn-in samples. During burn-in, we optimize the auxiliary variable sampling step size. Further hyper-parameters are described on \Cref{sec:SI-model}.

\textbf{Inferred values.} Our procedure gives posterior samples for each of $\theta_0, \theta_1, \alpha_\ell$ (and thus induced $\psi_i$), and $A_i$. We use these parameters to calculate $\Pr(A_i = 1 | \cdot)$.

\subsection{Framework overview} 

We use the above model as a foundation and present a novel framework for resource allocation in the presence of under-reporting. Our framework for evaluation and application, implemented in the remainder of the paper, is as follows.

First, we evaluate the approach \textit{semi-}synthetically using the real spatial map, showing model expressivity (that it can generate the full range of storm data), parameter recovery (that estimated parameter point estimates are correct and posteriors are calibrated), and predictive performance (that it correctly identifies unreported events $A_i$). 

Second, we evaluate performance on real storm data, without needing external ground truth data. In particular, we show that, using data in the initial hours after the storm, the model predicts \textit{future} reports in the following days. 

Third, we demonstrate the approach's \textit{application} to more efficient and equitable resource allocation. We show that, using the model predictions of ground truth unreported events $P(A_i)$, the resulting allocation better matches the population distribution and in particular does not deprioritize populations with lower reporting rates, unlike other approaches.

Finally, we \textit{pool} parameters across storms (using a Bayes factor approach as described in \Cref{sec:bayes}) to robustly characterize under-reporting behavior over time. Leveraging estimates of under-reporting behavior across storms would further aid in proactive resource allocation for future storms.  

\section{Data and models}

\paragraph{Data.} We apply our model to flood reports in New York City. We primarily use Census tracts as nodes in our graph; two tracts are adjacent if they share a border. Minimum-distance edges are drawn whenever needed to make the graph connected (e.g. linking Staten Island to Brooklyn). For node-specific features $X$, we use demographic variables obtained from Census data, which include population, socioeconomic status, and racial composition measurements.

We use 311 resident reports of street floods. For our main results, we look at reports in the week of September 1st through September 8th, following Hurricane Ida. The 311 reports dataset is publicly available in the NYC Open Data portal, supporting replication of our results. We split the data into train and test reporting periods: we fit our models with reports created until $8\%$ of the census tracts have received at least one report---this threshold is reached around 4 hours after the storm starting time. The remaining reports are used for evaluation. Out of the 2221 census tracts in our network, 177 tracts report at least one flooding incident during the training period, and 346 tracts report during the test period.

When evaluating historical under-reporting, we also consider reports of other floods in New York City. We look at data following the passage of Hurricane Henri (August 2021) and Tropical Storm Ophelia (September 2023). Further details on reports and data are presented in \Cref{sec:SI-data}.

\paragraph{Models evaluated.} We present results from four models: (a) Our model with \textit{homogeneous reporting}; (b) Our model with \textit{heterogeneous reporting}, with reporting probability varying as a function of demographic features; (c) a \emph{spatial baseline} model in which we predict test-time reporting as the fraction of a node's neighbors that reported during training period; (d) a \emph{Gaussian Process} (GP) baseline model, a standard approach to leverage geographic information. The baselines reflect approaches to incorporating spatial correlation, without explicitly modeling the reporting process as distinct from the incident occurrence process. Other modeling approaches (such as graph neural networks) may also be appropriate and perform well in terms of prediction, but it may be challenging to use such models to separately recover reporting from ground truth processes -- we leave such approaches to future work.

\section{Semi-synthetic simulation experiments}
\label{sec:simulation}

We verify that our models correctly recover the true parameters and latent states in semi-synthetic data settings where the ground truth parameters and latent state are known. To generate semi-synthetic data, we begin with the real NYC spatial network $E$ and demographic features $X$, and then for each of the two reporting models (homogeneous and heterogeneous), we generate latent states $A$ and reports $T$ through MCMC sampling assuming the corresponding data generating process. The observed data given to each model is $T$, $E$, and (for the heterogeneous reporting model) $X$. For all experiments, 500 trials were performed; for each trial, we re-sample new values of the latent parameters, re-generate $A$ and $T$, and run two MCMC chains for inference. Here, we report results when data is drawn according to the heterogeneous model; details and other results, including validating model expressivity, are in \Cref{sec:SI-synthetic}.  

\paragraph{Calibration and Identifiability.}

We verify that our inference procedure correctly recovers the true data generating parameters, a standard identifiability check~\citep{chang2021mobility,pierson2019inferring}. We find an overall high correlation between the recovered parameters and the true, latent parameter values. Correlation values all $0.60$. For the regression slope coefficients $\alpha_\ell$, which we more directly analyze, correlations of $0.91$ and higher are observed. We also verify that our confidence intervals are \emph{calibrated}, another standard check \cite{wilder_tracking_2021}: at each significance level, whether posterior distribution confidence intervals cover the correct fraction of true values.

\paragraph{Predictive Performance.}
An advantage of semi-synthetic data (in contrast to real data) is that the ground truth latent states $A_i$ are known. We can therefore compare the model's inferred event probabilities $\Pr(A_i)$ to the true latent states $A_i$. \Cref{table:AUC_synthetic} reports model AUC (area under the ROC curve), comparing the heterogeneous reporting model to the baselines and the homogeneous model \emph{when data is drawn following the heterogeneous model}. We find that correctly accounting for heterogeneous under-reporting has strong predictive performance, increasing increases AUC by $0.122$ from the homogeneous reporting model, by $0.129$ from the spatial baseline, and by $0.142$ from the GP baseline---all at a significance level of $10^{-4}$ or less. These values are repeated alongside analogous results for RMSE in Appendix \Cref{table:SI-performance_synthetic_heterogeneous}.

Overall, our semi-synthetic experiments validate that our model correctly recovers the true latent parameters, including how demographic covariates influence reporting rates. In this way, the use of spatial correlation overcomes the PU learning identifiability challenge. Further, our model outperforms baselines in the ability to infer the unobserved ground truth states $A_i$. Finally, the simulations demonstrate that, when true reporting processes are heterogeneous, assuming homogeneous under-reporting worsens estimation.

\begin{table}[tb]
	\centering
	\begin{tabular}{ c|c|c }
		\textbf{Model} & \textbf{AUC} & \textbf{95\% CI}\\
		\hline
		Heterogeneous Reporting & 0.642 & (0.637, 0.647) \\
		Homogeneous Reporting & 0.520 & (0.515, 0.525) \\
		GP Baseline & 0.501 & (0.497, 0.505) \\
		Spatial Baseline & 0.513 & (0.509, 0.518)\\
	\end{tabular}
	\caption
	{In simulation, average AUC to predict latent ground-truth $A_i$ according to each model. Nodes with observed training reports are excluded, as they are perfectly predicted by all models by definition. Confidence intervals were obtained through bootstrapping with 10,000 iterates.}
	\label{table:AUC_synthetic}
\end{table}

\section{Empirical Results}
\label{sec:results}

We now apply our framework to NYC 311 data. First, we fit the models to training report data from Hurricane Ida and show that our models outperform the baseline in terms of \textit{efficiency}, i.e., predicting future reports. Then, we show that accounting for heterogeneous under-reporting leads to \textit{more equitable} allocation of resources. Finally, looking at results across multiple storms, we investigate the socioeconomic and demographic features that are mostly associated with heterogeneity in under-reporting.

\paragraph{Prediction of floods and future reports.}

\begin{table*}[tb!]
	\centering
	\begin{subtable}[t]{\textwidth}
		\centering
		\begin{tabular}{c|c|c|c|c}
			\textbf{Model} & \textbf{AUC Estimate} & \textbf{AUC 95\% CI} & \textbf{RMSE Estimate} & \textbf{RMSE 95\% CI}\\
			\hline
			Heterogeneous Reporting & 0.680 & (0.646, 0.713) & 0.355 & (0.338, 0.371) \\
			Homogeneous Reporting & 0.682 & (0.649, 0.714) & 0.360 & (0.343, 0.376)\\
			GP Baseline & 0.629 & (0.595, 0.662) & 0.417 & (0.400, 0.434)\\
			Spatial Baseline & 0.647 & (0.616, 0.678) & 0.395 & (0.377, 0.412)\\
		\end{tabular}
		\caption{AUC and RMSE point estimates and confidence intervals for each of the four models.}
	\end{subtable}
	\newline
	\newline
	\begin{subtable}[t]{\textwidth}
		\centering
		\begin{tabular}{c|c|c|c|c|c}
			\textbf{Model A}& \textbf{Model B}& \textbf{$\Delta_{\text{AUC}}$} & \textbf{$\Delta_{\text{AUC}}$ p-value} & \textbf{$\Delta_{\text{RMSE}}$} & \textbf{$\Delta_{\text{RMSE}}$ p-value}\\
			\hline
			Heterogeneous Reporting & Homogeneous Reporting & -0.002 & 0.831 & -0.004 & 0.004\\
			& GP Baseline & 0.051 & 0.003 & -0.062 & $< 10^{-3}$\\
			& Spatial Baseline & 0.033 & 0.009 & -0.040 & $< 10^{-3}$\\
			\hline
			Homogeneous Reporting & GP Baseline & 0.054 & $< 10^{-3}$ & -0.058 & $< 10^{-3}$\\
			& Spatial Baseline & 0.035 & $< 10^{-3}$ & -0.035 & $< 10^{-3}$\\
		\end{tabular}
		\caption{AUC and RMSE changes. Two-sided p-values report whether model A and model B differ significantly in performance.}
	\end{subtable}
	\caption{Performance metrics for the four models in predicting future reports. Confidence intervals were obtained by bootstrapping the tracts with 10,000 iterates.}
	\label{table:MAIN_performance_Ida}
\end{table*}

\begin{figure}[tb!]
	\includegraphics[width=\columnwidth]{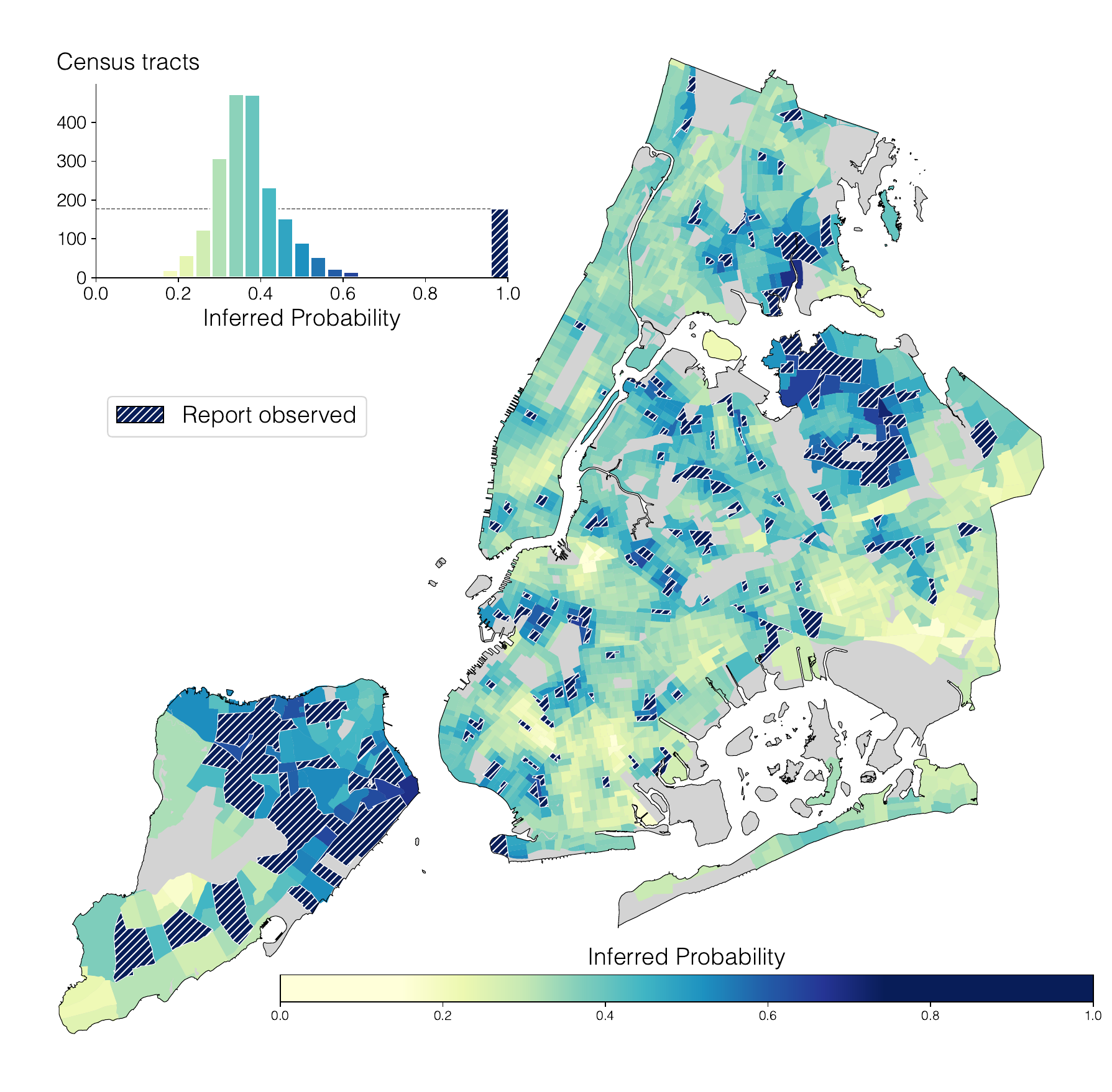}
	\caption{Model-inferred probabilities $\Pr(A_i)$ that each New York City Census tract is flooded after Hurricane Ida, from the heterogeneous reporting model. Hatched lines indicate tracts that reported during the training period.}\label{fig:MAIN-Aprob}
\end{figure}

\begin{figure}[tb!]
	\includegraphics[width=\columnwidth]{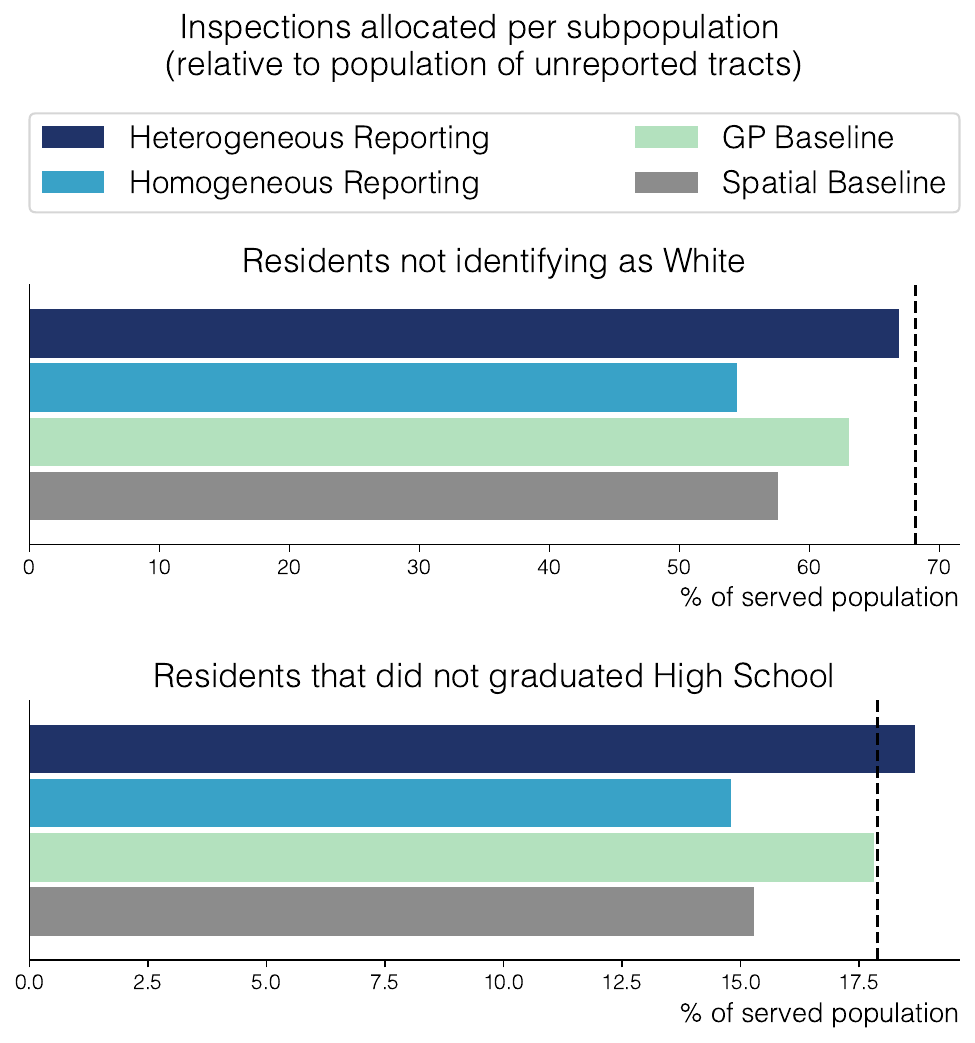}
	\caption
	{Demographic disparities when allocating resources to 100 census tracts (among those that do not report), using inferred flood probabilities from the four models. The horizontal axes shows the proportion of all residents served by the inspections (i.e. those who reside in the $100$ inspected census tracts) who are non-white and do not have a high school degree, computed as a weighted average from the proportions on inspected tracts. Dashed lines represent the total proportion of residents in tracts without a report who are non-white and do not have a high school degree.}
	\label{fig:MAIN-equity}
\end{figure}

\Cref{fig:MAIN-Aprob} shows, for the \textit{heterogeneous reporting model}, the inferred probability of flood by census tract, $\Pr(A_i)$. This map indicates that there is a substantial spatial correlation in reporting, but also that there is likely substantial under-reporting: many tracts have several neighboring tracts with reports, but did not report themselves. The positive spatial correlation is captured by our positive estimate for the spatial correlation parameter $\theta_1$ ($0.15$, 95\% CI ($0.08$, $0.21$)). Convergence diagnostics indicate that the inference procedure converged, with maximum $\hat R = 1.03$ for the latent parameters.

We evaluate the four models by how well they predict \textit{future reports.} Unlike the simulation results discussed in \Cref{sec:simulation}, we cannot evaluate the models in terms of predicting ground truth $A_i$: we do not have access to \textit{true} flooding events, just reports -- lack of such ground truth data indeed motivates the city to use 311 reporting data. We fit the model using train time data and then compare the model estimates of $\Pr(T)$ to future reports $T$ during the test period. This prediction task is decision-relevant because better prediction of future reports would allow the agency to proactively allocate resources (though, as we discuss below, the agency should still be aware of heterogeneous under-reporting).

\Cref{table:MAIN_performance_Ida} reports AUC and RMSE for each model, along with bootstrapped $95\%$  confidence intervals and relative improvements. The models accounting for under-reporting achieve better predictive performance than the baseline models that do not. The p-values shown test for positive \textit{differences} between each pair of models. Performance improvements are statistically significant at the 0.05 level for each of our models over the baselines. In practice, an improvement in report prediction -- when converted to estimates of ground truth $A_i$ -- would translate to more efficient resource allocation, as city governments can anticipate what areas will need attention after a disaster.

\paragraph{Equitable inspection allocation.}

Consider an agency that is allocating resources (such as emergency response, maintenance, or inspections) after storms, in response to reports. \Cref{table:MAIN_performance_Ida} suggests that our models could lead to more \textit{efficient} allocations, as they are more predictive of future reports after the first day. Here, we analyze how \textit{equitable} these allocations are, under each of the three models we study. We consider the task of allocating a fixed number of resources to Census tracts without reports ($T_i=0$) -- as most tracts receive no report, the agency can allocate some resources to such tracts. We suppose that the agency first infers flood probabilities $\Pr(A_i)$ for each tract in which no report was received, $T_i=0$. It then allocates resources to the $k$ tracts with the highest inferred probabilities $\Pr(A_i)$. 

\Cref{fig:MAIN-equity} shows the fraction of resources allocated to non-white residents and residents without a high school degree, when $k=100$, alongside the population fractions of the tracts without a report. The model accounting for heterogeneous reporting allocates resources more in line with the population distribution, especially in comparison to the homogeneous reporting model which accounts for under-reporting but not \textit{differences} between populations. Note that the model does so while achieving similar predictive performance, as established above. 

Additional results in \Cref{sec:SI-equity} show these results are robust to other values of $k$ and for other socioeconomic and demographic factors. The results establish that taking into account heterogeneous under-reporting leads to less deprioritization of non-white and socioeconomically disadvantaged populations (more in line with population distributions).

\paragraph{Socioeconomic factors of heterogeneous reporting.} 
\Cref{fig:MAIN-equity} suggests that the heterogenous reporting model identifies and corrects for demographic disparities in under-reporting. We now analyze such differences directly. To understand \textit{persistent} reporting behavior across storms, we also run our model for Hurricanes Ophelia (September 2023) and Henri (August 2021) and pool together the feature coefficients. The pooling method and (qualitatively identical) results for individual storms are in \Cref{sec:SI-storms}. 

\begin{figure}[h]
	\includegraphics[width=\columnwidth]{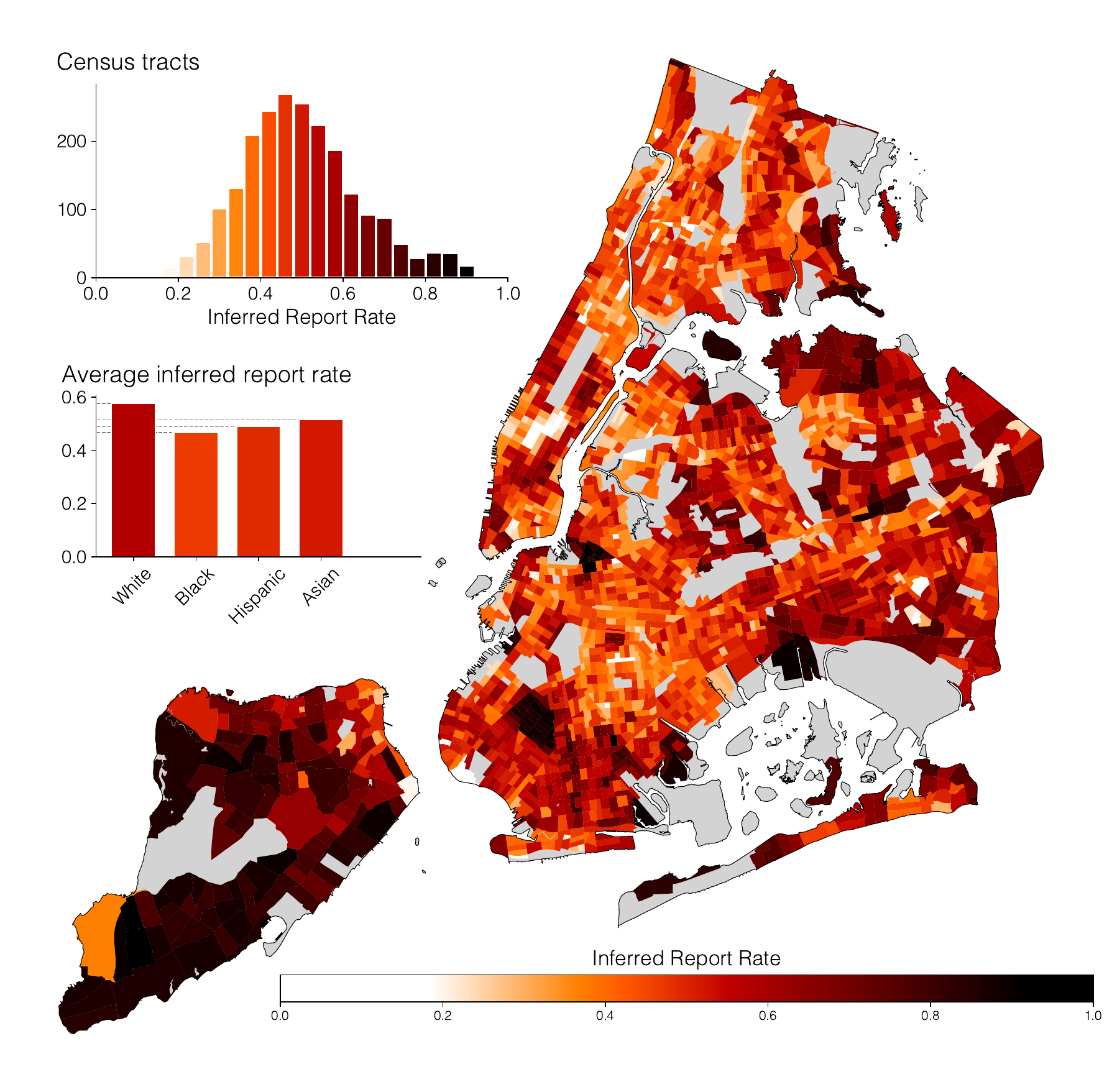}
	\caption{Model-inferred report rates $\psi_i$ per census tract, from the heterogeneous reporting model. The report rates range from near $0.1$ to $0.9$. Weighted averages of report rates per racial composition are shown in a bar plot.}\label{fig:MAIN-psi}
\end{figure}

Using the pooled regression coefficients, we estimate report rates $\psi_i$ for each census tract, with results shown in \Cref{fig:MAIN-psi}. Demographic patterns emerge -- e.g., the Upper West and Upper East Side neighborhoods in Manhattan, with higher median incomes, are estimated to have a higher reporting rate $\psi_i$ than surrounding areas, even though few reports were received in those areas. In contrast, the Bronx (relatively lower incomes) is estimated to have a low reporting rate. Taking a weighted average across neighborhoods, the reporting rate for white populations is, on average: $24\%$ higher than Black populations, $18\%$ higher than Hispanic populations, and $12\%$ higher than Asian populations.

\begin{figure}[h]
	\includegraphics[width=\columnwidth]{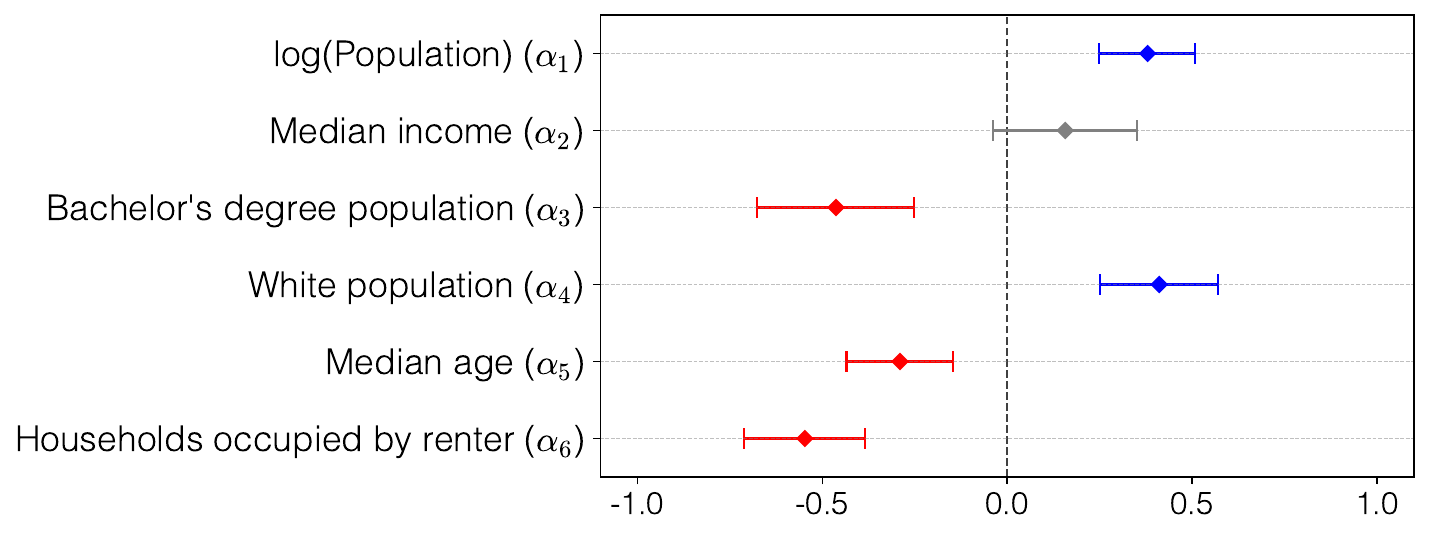}
	\caption{Estimated multivariate coefficients after pooling the three storms. Features were all standardized. Confidence intervals shown, and estimates with insignificant non-zero association colored in grey.}
	\label{fig:MAIN-multivariates}
\end{figure}

Next, we ask: what demographic factors are associated with under-reporting? \Cref{fig:MAIN-multivariates} shows the pooled posteriors for each coefficient $\alpha_\ell$ in equation \ref{eq:psi}: population, income, education, race/ethnicity, age, and household ownership.

We find that there are significant demographic disparities in reporting. As expected, a higher population is associated with higher reporting rates. However, other demographic factors are also associated with reporting rates. Higher proportions of white residents are positively correlated with reporting rate even when controlling for the other five demographic features considered, consistent with the different average report rates per subpopulation shown in \Cref{fig:MAIN-psi}. Median age and fraction of households occupied by a renter are negatively correlated with higher reporting, suggesting that neighborhoods with older, home-owning populations tend to receive reports at a higher rate. These estimates are consistent with prior work on demographic disparities in reporting in the NYC 311 system \cite{liu_quantifying_2023}. These estimates further explain the results in \Cref{fig:MAIN-equity}: the heterogeneous reporting model can identify and correct for these reporting disparities when calculating $P(A_i | \cdot)$, the probability that an area is actually flooded (even when no one submitted a report).

Further discussion, as well as similar analyses for other features, is in \Cref{sec:SI-univariates}. 

\FloatBarrier
\section{Discussion}
\label{sec:discussion}

This work shows the promise of leveraging the \textit{spatial correlation} of incidents to quantify and correct for \textit{under-reporting} in city crowdsourcing systems. Using a case study of flood reporting in New York City, we show that a Bayesian spatial model can accurately recover ground truth under-reporting patterns on semi-synthetic data by leveraging spatial correlation---a challenging task in missing data (PU learning) settings like the one we study. Using this model, we develop a framework for more accurate prediction of future flood reports compared to baselines, and allocation of resources that are more in line with population proportions (not neglecting neighborhoods with larger proportions of non-white and lower-income residents) -- improving both efficiency and equity. 

Future work might extend the model in several directions. (1) First, we fit the model on binarized reporting data: i.e., whether a location has \emph{any reports}, as opposed to the \emph{count} of reports. While, in our setting, a binary variable is sufficient to capture most of the variation (96.5\% of locations have zero or one report) the modeling approach here could plausibly be extended to accommodate count data as well. (2) Second, we assume that flooding is spatially correlated according to an Ising model, and do not explicitly model shared infrastructure and weather patterns. We adopt this model given the well-studied nature of Ising models, which aids estimation. Future work -- as driven by the model setting -- may choose to modify these choices, though model identifiability may be a challenge. (3) Third, in crowdsourcing settings, we can sometimes incorporate external data, such as (potentially noisy or non-randomly distributed) sensors. That is: if we have access to data from a different source (e.g. sensors to detect floods that are placed sparsely through the city) which reveals that $A_i = 1$ even though $T_i=0$ for some $i$, will estimates improve? This question is practically important: often inspections occur in clusters of regions, and it would be important for decision-makers to be able to incorporate the results of negative inspections in the model. (4) Fourth, 311 data is richer than data in other under-reporting settings (such as in ecology) due to spatiotemporal correlations in reporting rates across different incident types or different events of the same type. For example, report rates for flood events may be related to report rates for pest infestations or noise complaints, as both may be related to socioeconomic or other factors. We further note that we can use the pooled estimates of reporting rates across storms when new storms occur, leading to faster, more efficient and equitable allocation. All of these directions represent important opportunities for future work. 

\section{Acknowledgments}
The authors thank Sidhika Balachandar, Serina Chang, Zhi Liu, Allison Koenecke, Matt Franchi, and Arkadiy Saakyan for feedback. This research was supported by a Meta research award, Google Research Scholar award, a Cornell Tech Urban Tech grant, NSF CAREER \#2142419, a CIFAR Azrieli Global scholarship, a LinkedIn Research Award, and the Abby Joseph Cohen Faculty Fund.

\FloatBarrier
\bibliography{network_underreporting}

\begin{thebibliography}{52}
\providecommand{\natexlab}[1]{#1}

\bibitem[{Agonafir et~al.(2022)Agonafir, Pabon, Lakhankar, Khanbilvardi, and
  Devineni}]{agonafir2022understanding}
Agonafir, C.; Pabon, A.~R.; Lakhankar, T.; Khanbilvardi, R.; and Devineni, N.
  2022.
\newblock Understanding New York City street flooding through 311 complaints.
\newblock \emph{Journal of Hydrology}, 605: 127300.

\bibitem[{Arnold, Dobbie, and Hull(2022)}]{arnold2022measuring}
Arnold, D.; Dobbie, W.; and Hull, P. 2022.
\newblock Measuring racial discrimination in bail decisions.
\newblock \emph{American Economic Review}, 112(9): 2992--3038.

\bibitem[{Balachandar, Garg, and Pierson(2023)}]{sidhika2023selectivelabels}
Balachandar, S.; Garg, N.; and Pierson, E. 2023.
\newblock Domain constraints improve risk prediction when outcome data is
  missing.
\newblock \emph{{NeurIPS ML4H Symposium}}.

\bibitem[{Bekker and Davis(2020)}]{bekker_learning_2020}
Bekker, J.; and Davis, J. 2020.
\newblock Learning from positive and unlabeled data: a survey.
\newblock \emph{Machine Learning}, 109(4): 719--760.

\bibitem[{Besag(1974)}]{besag_spatial_1974}
Besag, J. 1974.
\newblock Spatial {Interaction} and the {Statistical} {Analysis} of {Lattice}
  {Systems}.
\newblock \emph{Journal of the Royal Statistical Society: Series B
  (Methodological)}, 36(2): 192--225.

\bibitem[{Cai et~al.(2020)Cai, Gaebler, Garg, and Goel}]{cai2020fair}
Cai, W.; Gaebler, J.; Garg, N.; and Goel, S. 2020.
\newblock Fair allocation through selective information acquisition.
\newblock In \emph{Proceedings of the AAAI/ACM Conference on AI, Ethics, and
  Society}, 22--28.

\bibitem[{Chang et~al.(2021)Chang, Pierson, Koh, Gerardin, Redbird, Grusky, and
  Leskovec}]{chang2021mobility}
Chang, S.; Pierson, E.; Koh, P.~W.; Gerardin, J.; Redbird, B.; Grusky, D.; and
  Leskovec, J. 2021.
\newblock Mobility network models of COVID-19 explain inequities and inform
  reopening.
\newblock \emph{Nature}, 589(7840): 82--87.

\bibitem[{{City of New York}(2023)}]{city_of_new_york_severe_2023}
{City of New York}. 2023.
\newblock Severe {Weather}.

\bibitem[{Cooper et~al.(2000)Cooper, Dyer, Frieze, and
  Rue}]{cooper_mixing_2000}
Cooper, C.; Dyer, M.~E.; Frieze, A.~M.; and Rue, R. 2000.
\newblock Mixing properties of the {Swendsen}–{Wang} process on the complete
  graph and narrow grids.
\newblock \emph{Journal of Mathematical Physics}, 41(3): 1499--1527.

\bibitem[{Coston, Rambachan, and Chouldechova(2021)}]{coston2021characterizing}
Coston, A.; Rambachan, A.; and Chouldechova, A. 2021.
\newblock Characterizing fairness over the set of good models under selective
  labels.
\newblock In \emph{International Conference on Machine Learning}, 2144--2155.
  PMLR.

\bibitem[{Della~Rocca and Milanesi(2022)}]{della_rocca_new_2022}
Della~Rocca, F.; and Milanesi, P. 2022.
\newblock The {New} {Dominator} of the {World}: {Modeling} the {Global}
  {Distribution} of the {Japanese} {Beetle} under {Land} {Use} and {Climate}
  {Change} {Scenarios}.
\newblock \emph{Land}, 11(4): 567.
\newblock Number: 4 Publisher: Multidisciplinary Digital Publishing Institute.

\bibitem[{Elkan and Noto(2008)}]{elkan_learning_2008}
Elkan, C.; and Noto, K. 2008.
\newblock Learning classifiers from only positive and unlabeled data.
\newblock In \emph{Proceedings of the 14th {ACM} {SIGKDD} international
  conference on {Knowledge} discovery and data mining}, {KDD} '08, 213--220.
  New York, NY, USA: Association for Computing Machinery.
\newblock ISBN 978-1-60558-193-4.

\bibitem[{{Federal Emergency Management Agency}(2023)}]{FEMAlearningagenda}
{Federal Emergency Management Agency}. 2023.
\newblock Appendix 2 - Learning Agenda.
\newblock
  \url{https://www.fema.gov/about/strategic-plan/appendices/learning-agenda}.

\bibitem[{Franchi et~al.(2023)Franchi, Zamfirescu-Pereira, Ju, and
  Pierson}]{franchi2023detecting}
Franchi, M.; Zamfirescu-Pereira, J.; Ju, W.; and Pierson, E. 2023.
\newblock Detecting disparities in police deployments using dashcam data.
\newblock In \emph{Proceedings of the 2023 ACM Conference on Fairness,
  Accountability, and Transparency}, 534--544.

\bibitem[{Garg, Li, and Monachou(2021)}]{garg2021standardized}
Garg, N.; Li, H.; and Monachou, F. 2021.
\newblock Standardized tests and affirmative action: The role of bias and
  variance.
\newblock In \emph{Proceedings of the 2021 ACM Conference on Fairness,
  Accountability, and Transparency}, 261--261.

\bibitem[{Guerdan et~al.(2023)Guerdan, Coston, Holstein, and
  Wu}]{guerdan2023counterfactual}
Guerdan, L.; Coston, A.; Holstein, K.; and Wu, Z.~S. 2023.
\newblock Counterfactual Prediction Under Outcome Measurement Error.
\newblock In \emph{Proceedings of the 2023 ACM Conference on Fairness,
  Accountability, and Transparency}, 1584--1598.

\bibitem[{Heikkinen and Hogmander(1994)}]{heikkinen_fully_1994}
Heikkinen, J.; and Hogmander, H. 1994.
\newblock Fully {Bayesian} {Approach} to {Image} {Restoration} with an
  {Application} in {Biogeography}.
\newblock \emph{Journal of the Royal Statistical Society. Series C (Applied
  Statistics)}, 43(4): 569--582.
\newblock Publisher: [Wiley, Royal Statistical Society].

\bibitem[{Jung et~al.(2018)Jung, Corbett-Davies, Shroff, and
  Goel}]{jung2018omitted}
Jung, J.; Corbett-Davies, S.; Shroff, R.; and Goel, S. 2018.
\newblock Omitted and included variable bias in tests for disparate impact.
\newblock \emph{arXiv preprint arXiv:1809.05651}.

\bibitem[{Kleinberg et~al.(2018)Kleinberg, Lakkaraju, Leskovec, Ludwig, and
  Mullainathan}]{kleinberg2018human}
Kleinberg, J.; Lakkaraju, H.; Leskovec, J.; Ludwig, J.; and Mullainathan, S.
  2018.
\newblock Human decisions and machine predictions.
\newblock \emph{The quarterly journal of economics}, 133(1): 237--293.

\bibitem[{Kontokosta, Hong, and Korsberg(2017)}]{kontokosta_equity_2017}
Kontokosta, C.; Hong, B.; and Korsberg, K. 2017.
\newblock Equity in 311 {Reporting}: {Understanding} {Socio}-{Spatial}
  {Differentials} in the {Propensity} to {Complain}.
\newblock ArXiv:1710.02452 [cs].

\bibitem[{Lakkaraju et~al.(2017)Lakkaraju, Kleinberg, Leskovec, Ludwig, and
  Mullainathan}]{lakkaraju2017selective}
Lakkaraju, H.; Kleinberg, J.; Leskovec, J.; Ludwig, J.; and Mullainathan, S.
  2017.
\newblock The selective labels problem: Evaluating algorithmic predictions in
  the presence of unobservables.
\newblock In \emph{Proceedings of the 23rd ACM SIGKDD International Conference
  on Knowledge Discovery and Data Mining}, 275--284.

\bibitem[{Laufer, Pierson, and Garg(2022)}]{end_to_end_auditing}
Laufer, B.; Pierson, E.; and Garg, N. 2022.
\newblock End-to-end Auditing of Decision Pipelines.
\newblock In \emph{ICML Workshop on Responsible Decision-Making in Dynamic
  Environments. ACM, Baltimore, Maryland, USA}, 1--7.

\bibitem[{Liu et~al.(2003)Liu, Dai, Li, Lee, and Yu}]{liu_building_2003}
Liu, B.; Dai, Y.; Li, X.; Lee, W.; and Yu, P. 2003.
\newblock Building text classifiers using positive and unlabeled examples.
\newblock In \emph{Third {IEEE} {International} {Conference} on {Data}
  {Mining}}, 179--186.

\bibitem[{Liu, Bhandaram, and Garg(2023)}]{liu_quantifying_2023}
Liu, Z.; Bhandaram, U.; and Garg, N. 2023.
\newblock Quantifying spatial under-reporting disparities in resident
  crowdsourcing.
\newblock \emph{Nature Computational Science}.

\bibitem[{Liu, Rankin, and Garg(2024)}]{liu2023library}
Liu, Z.; Rankin, S.; and Garg, N. 2024.
\newblock Identifying and Addressing Disparities in Public Libraries\\ with
  Bayesian Latent Variable Modeling.
\newblock In \emph{Proceedings of the AAAI Conference on Artificial
  Intelligence}.

\bibitem[{Mauerman et~al.(2022)Mauerman, Tellman, Lall, Tedesco, Colosio,
  Thomas, Osgood, and Bhuyan}]{mauerman_high-quality_2022}
Mauerman, M.; Tellman, E.; Lall, U.; Tedesco, M.; Colosio, P.; Thomas, M.;
  Osgood, D.; and Bhuyan, A. 2022.
\newblock High-{Quality} {Historical} {Flood} {Data} {Reconstruction} in
  {Bangladesh} {Using} {Hidden} {Markov} {Models}.
\newblock In Tarekul~Islam, G.~M.; Shampa, S.; and Chowdhury, A. I.~A., eds.,
  \emph{Water {Management}: {A} {View} from {Multidisciplinary}
  {Perspectives}}, 191--210. Cham: Springer International Publishing.
\newblock ISBN 978-3-030-95722-3.

\bibitem[{Minkoff(2016)}]{minkoff_nyc_2016}
Minkoff, S.~L. 2016.
\newblock {NYC} 311: {A} {Tract}-{Level} {Analysis} of {Citizen}–{Government}
  {Contacting} in {New} {York} {City}.
\newblock \emph{Urban Affairs Review}, 52(2): 211--246.
\newblock Publisher: SAGE Publications Inc.

\bibitem[{Mosavi, Ozturk, and Chau(2018)}]{mosavi2018flood}
Mosavi, A.; Ozturk, P.; and Chau, K.-w. 2018.
\newblock Flood prediction using machine learning models: Literature review.
\newblock \emph{Water}, 10(11): 1536.

\bibitem[{Movva et~al.(2023)Movva, Shanmugam, Hou, Pathak, Guttag, Garg, and
  Pierson}]{movva2023coarse}
Movva, R.; Shanmugam, D.; Hou, K.; Pathak, P.; Guttag, J.; Garg, N.; and
  Pierson, E. 2023.
\newblock Coarse race data conceals disparities in clinical risk score
  performance.
\newblock \emph{arXiv preprint arXiv:2304.09270}.

\bibitem[{Murray, Ghahramani, and MacKay(2006)}]{murray_mcmc_nodate}
Murray, I.; Ghahramani, Z.; and MacKay, D. J.~C. 2006.
\newblock MCMC for Doubly-Intractable Distributions.
\newblock In \emph{Proceedings of the Twenty-Second Conference on Uncertainty
  in Artificial Intelligence}, UAI'06, 359–366. Arlington, Virginia, USA:
  AUAI Press.
\newblock ISBN 0974903922.

\bibitem[{Møller et~al.(2006)Møller, Pettitt, Reeves, and
  Berthelsen}]{moller_efficient_2006}
Møller, J.; Pettitt, A.~N.; Reeves, R.; and Berthelsen, K.~K. 2006.
\newblock An {Efficient} {Markov} {Chain} {Monte} {Carlo} {Method} for
  {Distributions} with {Intractable} {Normalising} {Constants}.
\newblock \emph{Biometrika}, 93(2): 451--458.
\newblock Publisher: [Oxford University Press, Biometrika Trust].

\bibitem[{Newman(2021)}]{newman_43_2021}
Newman, A. 2021.
\newblock 43 {Die} as {Deadliest} {Storm} {Since} {Sandy} {Devastates} the
  {Northeast}.
\newblock \emph{The New York Times}.

\bibitem[{{NYC Open Data}(2023)}]{nyc_open_data_311_nodate}
{NYC Open Data}. 2023.
\newblock 311 {Service} {Requests} from 2010 to {Present}.

\bibitem[{Obermeyer et~al.(2019)Obermeyer, Powers, Vogeli, and
  Mullainathan}]{obermeyer2019dissecting}
Obermeyer, Z.; Powers, B.; Vogeli, C.; and Mullainathan, S. 2019.
\newblock Dissecting racial bias in an algorithm used to manage the health of
  populations.
\newblock \emph{Science}, 366(6464): 447--453.

\bibitem[{Onsager(1944)}]{onsager_crystal_1944}
Onsager, L. 1944.
\newblock Crystal {Statistics}. {I}. {A} {Two}-{Dimensional} {Model} with an
  {Order}-{Disorder} {Transition}.
\newblock \emph{Physical Review}, 65(3-4): 117--149.
\newblock Publisher: American Physical Society.

\bibitem[{O’Brien et~al.(2017)O’Brien, Offenhuber, Baldwin-Philippi, Sands,
  and Gordon}]{obrien_uncharted_2017}
O’Brien, D.~T.; Offenhuber, D.; Baldwin-Philippi, J.; Sands, M.; and Gordon,
  E. 2017.
\newblock Uncharted {Territoriality} in {Coproduction}: {The} {Motivations} for
  311 {Reporting}.
\newblock \emph{Journal of Public Administration Research and Theory}, 27(2):
  320--335.

\bibitem[{O’Brien, Sampson, and Winship(2015)}]{doi:10.1177/0081175015576601}
O’Brien, D.~T.; Sampson, R.~J.; and Winship, C. 2015.
\newblock Ecometrics in the Age of Big Data: Measuring and Assessing “Broken
  Windows” Using Large-scale Administrative Records.
\newblock \emph{Sociological Methodology}, 45(1): 101--147.

\bibitem[{Park et~al.(2017)Park, Jang, Galanis, Shin, Stefankovic, and
  Vigoda}]{park_rapid_2017}
Park, S.; Jang, Y.; Galanis, A.; Shin, J.; Stefankovic, D.; and Vigoda, E.
  2017.
\newblock Rapid {Mixing} {Swendsen}-{Wang} {Sampler} for {Stochastic}
  {Partitioned} {Attractive} {Models}.
\newblock ArXiv:1704.02232 [cs, stat].

\bibitem[{Pierson(2020)}]{pierson2020assessing}
Pierson, E. 2020.
\newblock Assessing racial inequality in COVID-19 testing with Bayesian
  threshold tests.
\newblock \emph{NeurIPS ML4H Workshop}.

\bibitem[{Pierson et~al.(2019)Pierson, Koh, Hashimoto, Koller, Leskovec,
  Eriksson, and Liang}]{pierson2019inferring}
Pierson, E.; Koh, P.~W.; Hashimoto, T.; Koller, D.; Leskovec, J.; Eriksson, N.;
  and Liang, P. 2019.
\newblock Inferring multidimensional rates of aging from cross-sectional data.
\newblock In \emph{The 22nd International Conference on Artificial Intelligence
  and Statistics}, 97--107. PMLR.

\bibitem[{Rambachan et~al.(2021)}]{rambachan2021identifying}
Rambachan, A.; et~al. 2021.
\newblock Identifying prediction mistakes in observational data.
\newblock \emph{Harvard University}.

\bibitem[{Santos-Fernandez et~al.(2021)Santos-Fernandez, Peterson, Vercelloni,
  Rushworth, and Mengersen}]{santos-fernandez_correcting_2021}
Santos-Fernandez, E.; Peterson, E.~E.; Vercelloni, J.; Rushworth, E.; and
  Mengersen, K. 2021.
\newblock Correcting {Misclassification} {Errors} in {Crowdsourced}
  {Ecological} {Data}: {A} {Bayesian} {Perspective}.
\newblock \emph{Journal of the Royal Statistical Society Series C: Applied
  Statistics}, 70(1): 147--173.

\bibitem[{Shanmugam and Pierson(2021)}]{shanmugam2021quantifying}
Shanmugam, D.; and Pierson, E. 2021.
\newblock Quantifying Inequality in Underreported Medical Conditions.
\newblock \emph{arXiv preprint arXiv:2110.04133}.

\bibitem[{Sicacha-Parada et~al.(2021)Sicacha-Parada, Steinsland, Cretois, and
  Borgelt}]{sicacha-parada_accounting_2021}
Sicacha-Parada, J.; Steinsland, I.; Cretois, B.; and Borgelt, J. 2021.
\newblock Accounting for spatial varying sampling effort due to accessibility
  in {Citizen} {Science} data: {A} case study of moose in {Norway}.
\newblock \emph{Spatial Statistics}, 42: 100446.

\bibitem[{Spezia, Friel, and Gimona(2018)}]{spezia_spatial_2018}
Spezia, L.; Friel, N.; and Gimona, A. 2018.
\newblock Spatial hidden {Markov} models and species distributions.
\newblock \emph{Journal of Applied Statistics}, 45(9): 1595--1615.

\bibitem[{Suwardi et~al.(2015)Suwardi, Dharma, Satya, and Lestari}]{geohash}
Suwardi, I.~S.; Dharma, D.; Satya, D.~P.; and Lestari, D.~P. 2015.
\newblock Geohash index based spatial data model for corporate.
\newblock In \emph{2015 International Conference on Electrical Engineering and
  Informatics (ICEEI)}, 478--483.

\bibitem[{Swendsen and Wang(1987)}]{swendsen_nonuniversal_1987}
Swendsen, R.~H.; and Wang, J.-S. 1987.
\newblock Nonuniversal critical dynamics in {Monte} {Carlo} simulations.
\newblock \emph{Physical Review Letters}, 58(2): 86--88.
\newblock Publisher: American Physical Society.

\bibitem[{Wilder, Mina, and Tambe(2021)}]{wilder_tracking_2021}
Wilder, B.; Mina, M.; and Tambe, M. 2021.
\newblock Tracking {Disease} {Outbreaks} from {Sparse} {Data} with {Bayesian}
  {Inference}.
\newblock \emph{Proceedings of the AAAI Conference on Artificial Intelligence},
  35(6): 4883--4891.

\bibitem[{Wolff(1989)}]{wolff_collective_1989}
Wolff, U. 1989.
\newblock Collective {Monte} {Carlo} {Updating} for {Spin} {Systems}.
\newblock \emph{Physical Review Letters}, 62(4): 361--364.
\newblock Publisher: American Physical Society.

\bibitem[{Xu et~al.(2023)Xu, Rolf, Beery, Bennett, Berger-Wolf, Birch,
  Bondi-Kelly, Brashares, Chapman, Corso et~al.}]{xu2023reflections}
Xu, L.; Rolf, E.; Beery, S.; Bennett, J.~R.; Berger-Wolf, T.; Birch, T.;
  Bondi-Kelly, E.; Brashares, J.; Chapman, M.; Corso, A.; et~al. 2023.
\newblock Reflections from the Workshop on AI-Assisted Decision Making for
  Conservation.
\newblock \emph{arXiv preprint arXiv:2307.08774}.

\bibitem[{Zanger-Tishler, Nyarko, and Goel(2023)}]{zanger2023risk}
Zanger-Tishler, M.; Nyarko, J.; and Goel, S. 2023.
\newblock Risk scores, label bias, and everything but the kitchen sink.
\newblock \emph{arXiv preprint arXiv:2305.12638}.

\bibitem[{Zink, Obermeyer, and Pierson(2023)}]{zink2023race}
Zink, A.; Obermeyer, Z.; and Pierson, E. 2023.
\newblock Race Corrections in Clinical Models: Examining Family History and
  Cancer Risk.
\newblock \emph{medRxiv}, 2023--03.

\end{thebibliography}
\appendix
\onecolumn

\section{Model Results for Other Storms}
\label{sec:SI-storms}

We repeat our empirical process for other storms, and then \textit{pool} our estimates together across storms. In particular, we run our heterogeneous reporting model on two other New York City storms: Hurricane Henri (August 2021) and Tropical Storm Ophelia (September 2023). We compare our estimates across storms to each other, finding remarkable consistency. We further compare the results with those of \cite{liu_quantifying_2023}, who also estimate census tract-level reporting rates in NYC 311, though using a different method that leverages \textit{duplicate} reports about the same incident. 

\subsection{Method for Pooling results across storm events}
\label{sec:bayes}
A benefit of our Bayesian approach is that we can pool model estimates across storm events. While some parameters such as reporting intercepts ($\alpha_0$), overall flooding frequency ($\theta_0$), and spatial correlation ($\theta_1$) are likely to be storm specific, other reporting parameters $\alpha_\ell$ (representing effects of demographic or socioeconomic factors) may be relatively consistent across events. We thus pool our estimates across events to increase statistical power.

Pooling is done via a Bayesian model pooling approach. We assume that the data (reports for each node for a given storm) are conditionally independent across storms, given the model parameters and flood occurrence -- i.e., whether a node (or two separate nodes) make reports in two separate events is not correlated, given the ground truth (similar to the assumption within each storm). Given this assumption, we can pool the results as follows.

For a given parameter $\alpha$ (for example, the coefficient corresponding to the population density or median income), suppose we have posterior distributions for a set of storms $\{1, \dots k, \dots K\}$ -- i.e., given the report data $\vec{T}_k$ for each storm $k$, we have a posterior $\Pr(\alpha \given \vec{T}_k)$. We want to estimate the posterior of the parameter given \textit{all} the storm data, which is:
$$\Pr\left(\alpha \given \vec{T}_1, \dots \vec{T}_K\right)$$

We do so as follows:

\begin{align}
	\Pr\left(\alpha \given \vec{T}_1, \dots \vec{T}_K\right) &= \frac{\Pr\left(\vec{T}_1, \dots \vec{T}_K \given \alpha\right) \cdot \Pr(\alpha) }{\Pr\left(\vec{T}_1, \dots \vec{T}_K\right)} & \text{Bayes' rule} \nonumber \\
	&= \frac{\prod_{k=1}^K \left[\Pr\left(\vec{T}_k \given \alpha\right)\right]\cdot \Pr(\alpha)}{\Pr\left(\vec{T}_1, \dots \vec{T}_K\right)} & \text{conditional indep.} \nonumber\\
	&= \prod_{k=1}^K\left[\frac{\Pr\left(\alpha\given \vec{T}_k\right)\cdot \Pr\left(\vec{T}_k\right)}{\Pr(\alpha)}\right]\cdot \frac{\Pr(\alpha)}{\Pr\left(\vec{T}_1, \dots \vec{T}_K\right)} & \text{Bayes' rule} \nonumber\\
	&= C\cdot \prod_{k=1}^K\left[\frac{\Pr\left(\alpha\given \vec{T}_k\right)}{\Pr(\alpha)}\right]\cdot \Pr(\alpha),\quad\text{where } C=\frac{\prod_{k=1}^K\Pr\left(\vec{T}_k\right)}{\Pr\left(\vec{T}_1, \dots \vec{T}_K\right)} \label{eqn:finalpooled}
\end{align}
Where $\frac{\Pr(\alpha \given \vec{T}_k)}{\Pr(\alpha)}$ is called the \textit{Bayes factor} for a given storm $k$ and parameter $\alpha$. The data-dependent term $C$ is a normalization constant that can be computed so that the pooled posterior is a valid probability distribution integrating to 1. Note that our Bayesian procedure for each storm gives us \textit{samples} from the posterior $\Pr(\alpha \given \vec{T}_k)$ for each $k$, and $\Pr(\alpha)$ corresponds to the known prior. Thus, to get our pooled posterior estimates $\Pr(\alpha \given \vec{T}_k)$ for each parameter, we do the following procedure:
\begin{enumerate}
	\item Fit a normal distribution to the posterior samples for each storm.
	\item Calculate the probability density function at each point for the posterior using \Cref{eqn:finalpooled}, the above distributions, and the prior distribution.
	\item Calculate the posterior summary statistics (95\% confidence interval, mean, median) from the calculated probability density function.
\end{enumerate}
Note that, crucially, this procedure does not require refitting models after each storm -- we can simply leverage the posterior distributions estimated using each storm's reports separately. After a new storm, we can create an updated pooled estimate without refitting models from other storms. 

\subsection{Parameter Estimates for Multiple Storms}

\begin{figure}[ht]
	\centering
	\includegraphics[width=\textwidth]{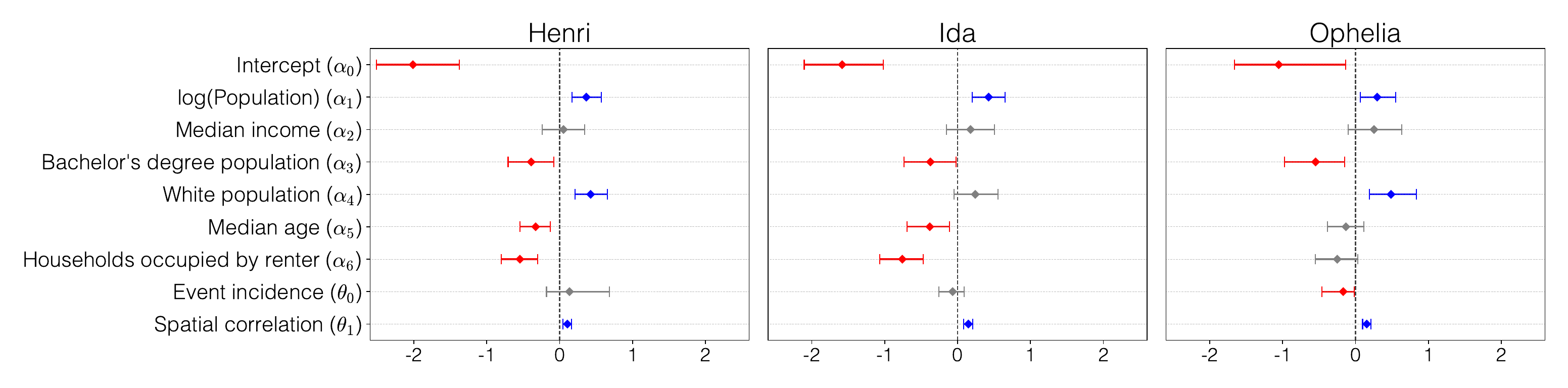}
	\caption
	{Estimated coefficients for the model trained in each storm individually. For the regression component, all features were standardized to have zero mean and unit variance. Confidence intervals are shown, and estimates with insignificant positive or negative associations are colored in grey.}
	\label{fig:SI-coefficients-multivariate-census}
\end{figure}

We showed in \Cref{fig:MAIN-multivariates} the pooled multivariate regression coefficients from our heterogeneous reporting model. We present results from the individual events---Hurricane Henri, Hurricane Ida, and Tropical Storm Ophelia---that were used to produce these estimates in \Cref{fig:SI-coefficients-multivariate-census}. Estimates for storm-specific parameters ($\theta_0$, $\theta_1$, and $\alpha_0$) were not pooled. The regression coefficient results, however, qualitatively agree across storms: the coefficients do not change direction, but may become statistically significant when pooled. We believe the pooled results are more powerful in providing an interpretation for historical demographic and socioeconomic trends of under-reporting.

\subsection{Tract Level Reporting Rates Across Multiple Storms}

\begin{figure}[ht!]
	\centering
	\includegraphics[width=\textwidth]{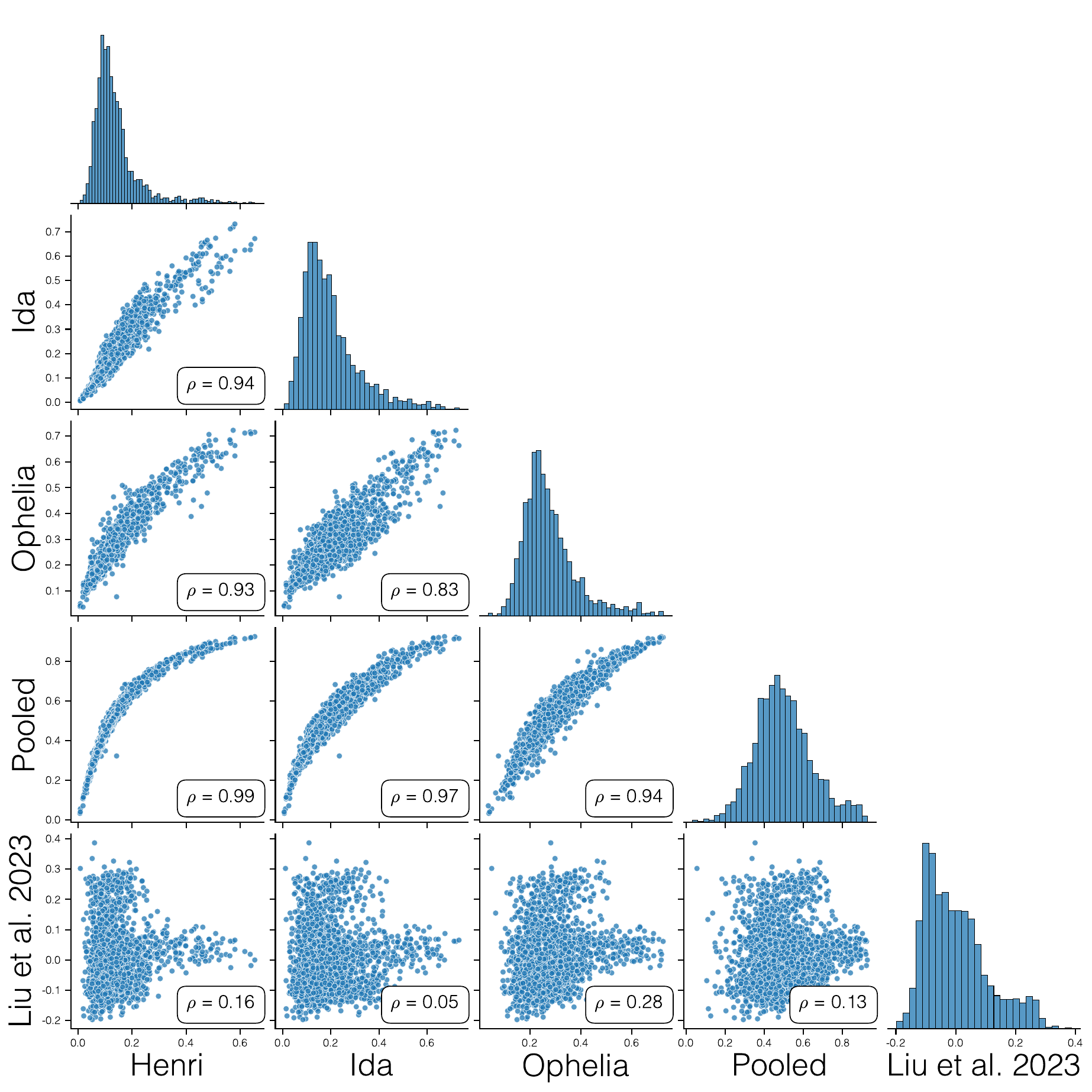}
	\caption
	{Correlation between the inferred report rates $\psi_i$ per \textbf{census tract }for the three different storms considered, as well as the pooled posteriors. The high rank correlation between results for all storms justifies using the pooled results in lieu of results for a specific storm to estimate the report rate. We also include correlations between our model and the model in \citet{liu_quantifying_2023}, previous work focusing on 311 reports of fallen trees.}
	\label{fig:SI-correlation-census}
\end{figure}

In the main text, we presented a map of inferred census tract report rates $\psi_i$ in Hurricane Ida (\Cref{fig:MAIN-psi}). We could infer $\psi_i$ similarly for any of the three storms. \Cref{fig:SI-correlation-census} shows the resulting scatter plots and correlation. We find a high internal correlation of our method's results across storms. Note that, to compute the pooled report rate, we ignore the intercept $\alpha_0$---whose role is simply to normalize for the average number of observed reports $\vec{T}$---i.e.

\begin{equation}\label{eq:SI-psi-pool}
	\psi^{\text{pool}}_i = \text{logit}^{-1}\left(\sum_{\ell=1}^M\alpha^{\text{pool}}_{\ell} X_{i\ell}\right),
\end{equation}

We also verify that our report rates correlate with previous work on 311 reporting behavior, namely the work of \citet{liu_quantifying_2023}.The correlation with these results is positive but weaker, reflecting the different model estimand and data (reporting rates for tree-related incidents), as well as methodological discrepancies (\citet{liu_quantifying_2023} use duplicate reports for a given incident). 

\clearpage
\FloatBarrier
\section{Model Results for Geohash Networks}
\label{sec:SI-geohash}

To verify our results are robust to different graphs (on the same underlying spatial setting), we re-run our models using geohashes. Geohashes encode latitude and longitudes as strings to partition the globe in rectangles of similar areas (see \citet{geohash} for some properties). We use resolution-6 geohashes and construct a network with $1238$ nodes after pre-processing to remove nodes with high water area and low population. Therefore, our geohash network has around $54\%$ as many nodes as the Census tract network used in our primary results. To associate demographics with each geohash, we project census tract demographic onto the geohash geometry---weighting each tract by the assumed population of the geohash that it contains.

In this section, we show that the model trained on geohashes is consistent (i) within itself, across different storms, and (ii) with the results presented in the main text for the census tract model.

\subsection{Consistency of the Geohash Model Across Multiple Storms}

\begin{figure}[ht]
	\centering
	\includegraphics[width=\textwidth]{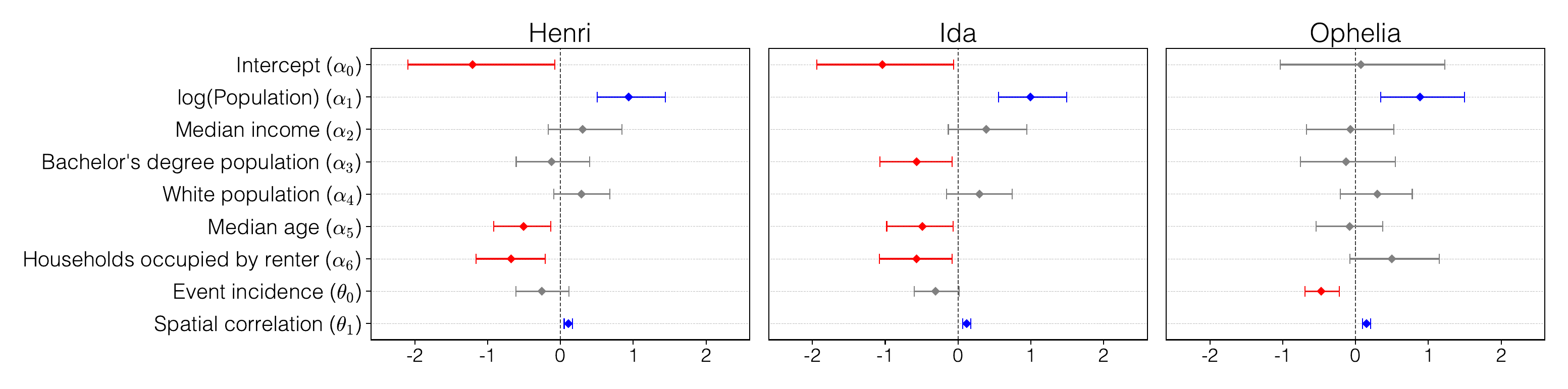}
	\caption
	{Estimated coefficients for the model trained in each storm individually withe the geohash network. For the regression component, all features were standardized to have zero mean and unit variance. Confidence intervals are shown, and estimates with insignificant positive or negative associations are colored in grey.}
	\label{fig:SI-coefficients-multivariate-geohash}
\end{figure}

We present results showing that the geohash model is consistent across the three storms considered, mirroring \Cref{sec:SI-storms}. First, \Cref{fig:SI-coefficients-multivariate-geohash} shows the parameter estimates inferred from the models. Similar to what we see in \Cref{fig:SI-coefficients-multivariate-census}, most of the regression coefficients $\alpha$ directionally agree across the three storms. We note, however, that the confidence intervals are wider in the case of geohash, resulting in many statistically insignificant coefficients. Increasing the sample size through using resolution-7 geohashes, for example, could address this issue, as there are nearly $50000$ nodes in such graph. However, this procedure would increase model run time and significantly sparsify the reports $\vec{T}$, affecting model priors and other hyperparameters. 

\begin{figure}[ht]
	\centering
	\includegraphics[width=\textwidth]{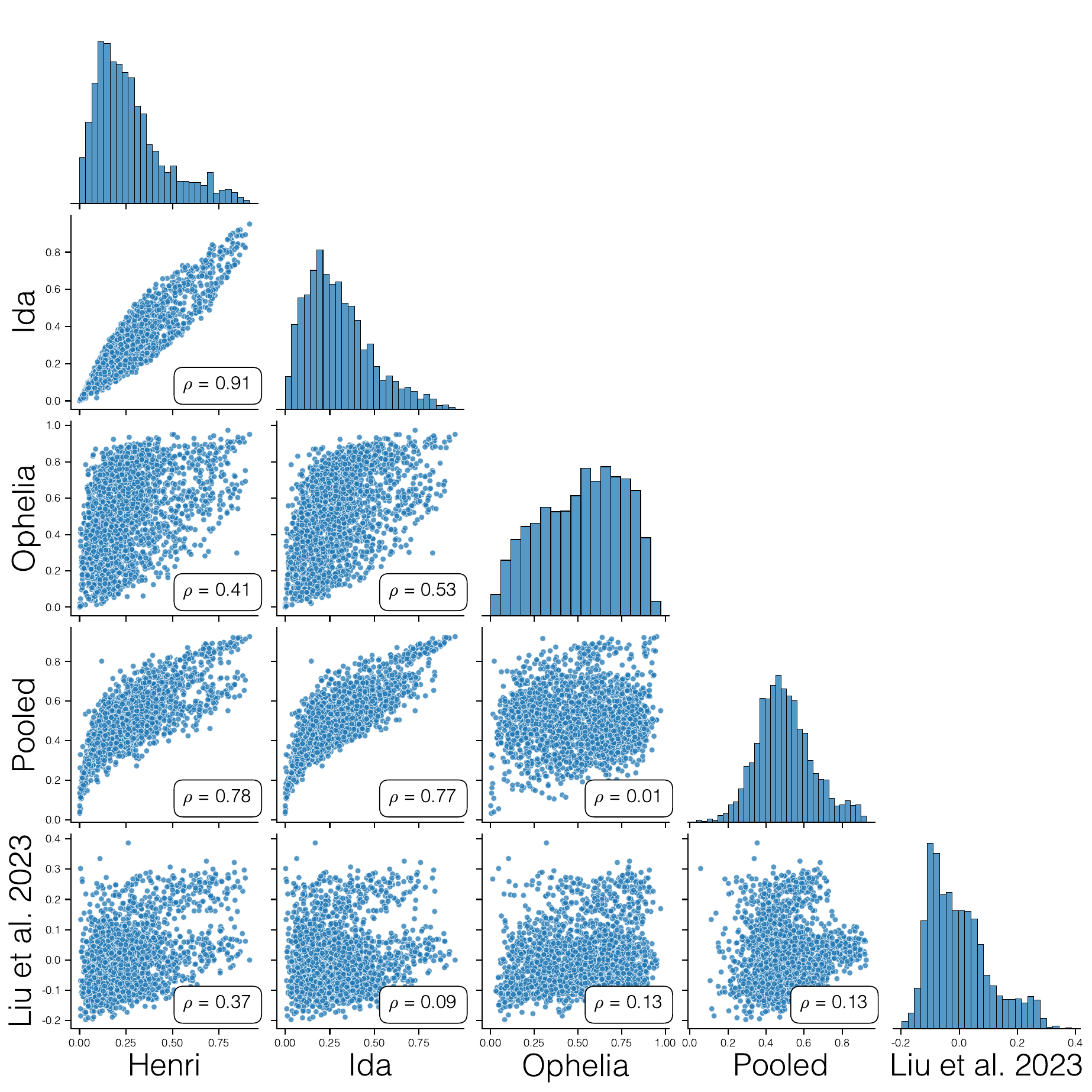}
	\caption
	{Correlation between the inferred report rates $\psi_i$ per \textbf{geohash} for the three different storms considered and the work of \citet{liu_quantifying_2023}. To compute the correlation with the latter, where coefficients are presented per census tract, we apply the estimates attained from geohash models to tract level demographic features.}
	\label{fig:SI-correlation-geohash}
\end{figure}

Second, we verify in \Cref{fig:SI-correlation-geohash} that the predicted report rates for different storms are correlated. We again find positive rank correlation across results for all three storms. We note that we find lower correlation when it comes to Tropical Storm Ophelia (2023). As shown in \Cref{fig:SI-coefficients-multivariate-geohash}, the regression coefficients for this storm are mostly non significant and four of them have point estimates very close to zero, so that we expect the inferred $\psi_i$ to have high uncertainty.

\subsection{Consistency of the Geohash and the Census Tract Models}

\begin{figure}[ht]
	\centering
	\includegraphics[width=\textwidth]{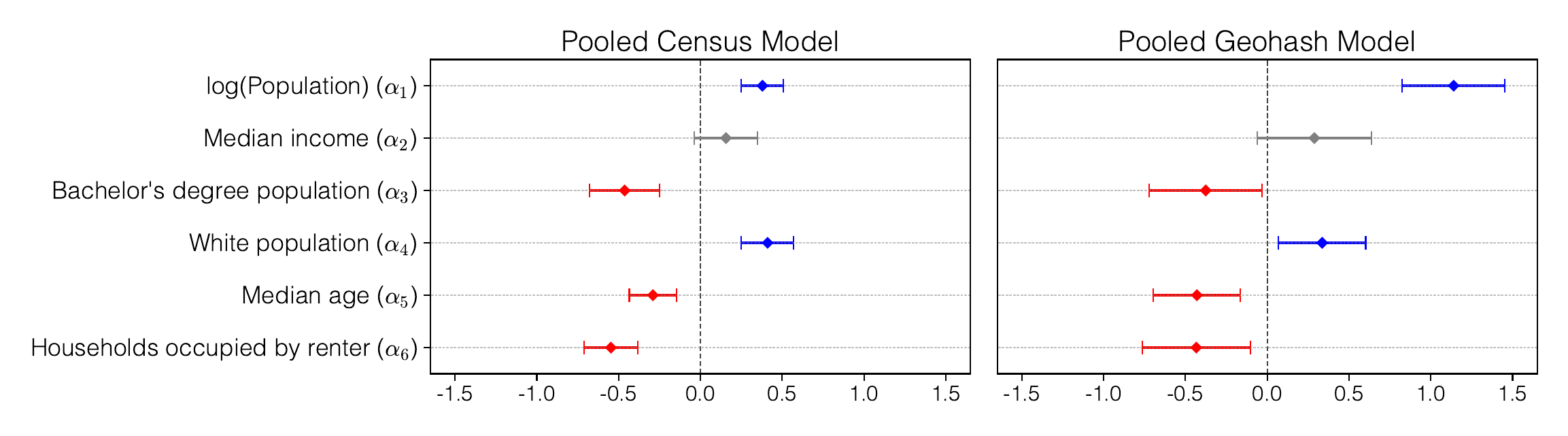}
	\caption
	{Pooled multivariate coefficients for census tract and geohash models.}
	\label{fig:SI-multivariates-census-and-geohash}
\end{figure}

\begin{figure}[ht]
	\centering
	\includegraphics[width=\textwidth]{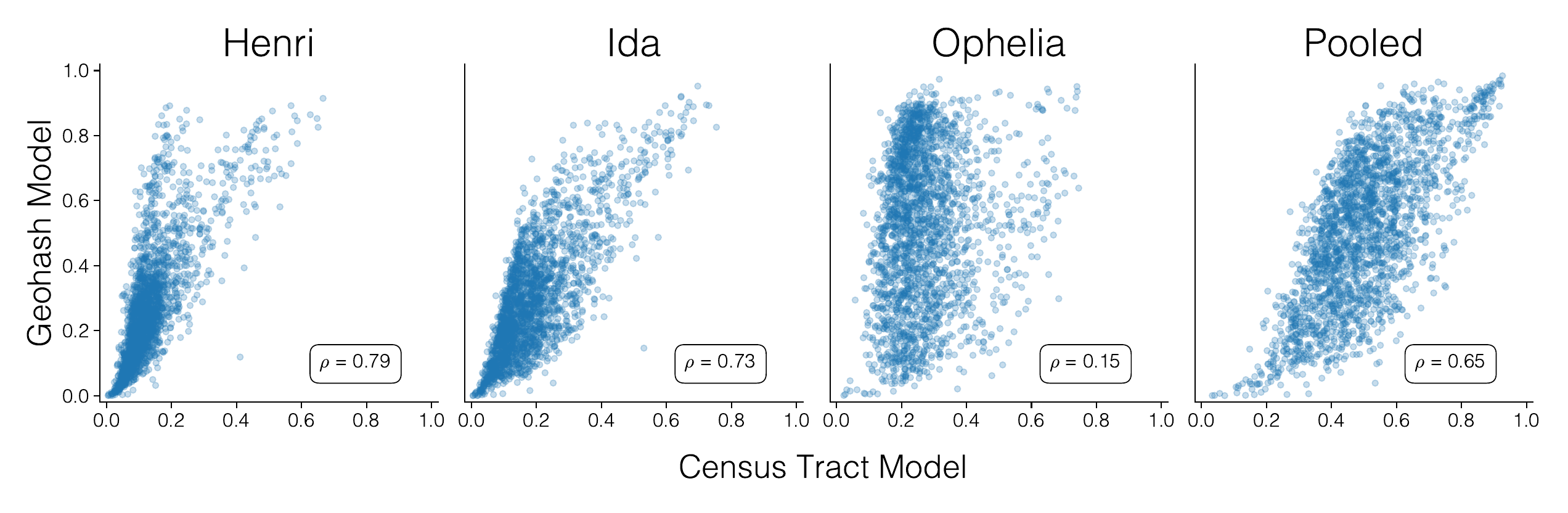}
	\caption
	{Inferred report rates $\psi_i$ per census tract according to the coefficients estimated by Geohash and Census Tract Models in each of the storms as well as for the pooled model.}
	\label{fig:SI-correlation-census-and-geohash}
\end{figure}

We further verify that the geohash models produce estimates consistent with the ones produced by the census tract models, reported on the main text. \Cref{fig:SI-multivariates-census-and-geohash} shows pooled multivariate estimates from geohashes along with main text pooled census estimates. All the coefficients agree directionally, although geohashes produce wider posteriors. Finally, \Cref{fig:SI-correlation-census-and-geohash} shows that, when we apply the multivariate estimates from each of the model to demographic features, the inferred report rates have positive correlation.

\clearpage
\FloatBarrier
\section{Univariate Heterogeneous Reporting Models}
\label{sec:SI-univariates}
In our primary analyses, we report the result of \textit{multivariate} reporting coefficients, i.e., include both population and income in the same model. A standard limitation of such an approach is that these coefficients have high co-linearity in NYC, limiting interpretation.  We further cannot use the findings from \Cref{fig:MAIN-multivariates} to report associations between each feature and the report probability. Additionally, many other features of interest were not included, so as to not to incur further co-linearity.

To address these issues, we train our heterogeneous reporting models with \emph{one socioeconomic or demographic feature at a time}. These univariate models present faster convergence than the full heterogeneous reporting model presented on the main text. We draw 15,000 samples from three chains, after 10,000 burn-in samples. All other sampling hyper-parameters remain the same---including the network.

\begin{figure}[ht]
	\centering
	\includegraphics[width=\textwidth]{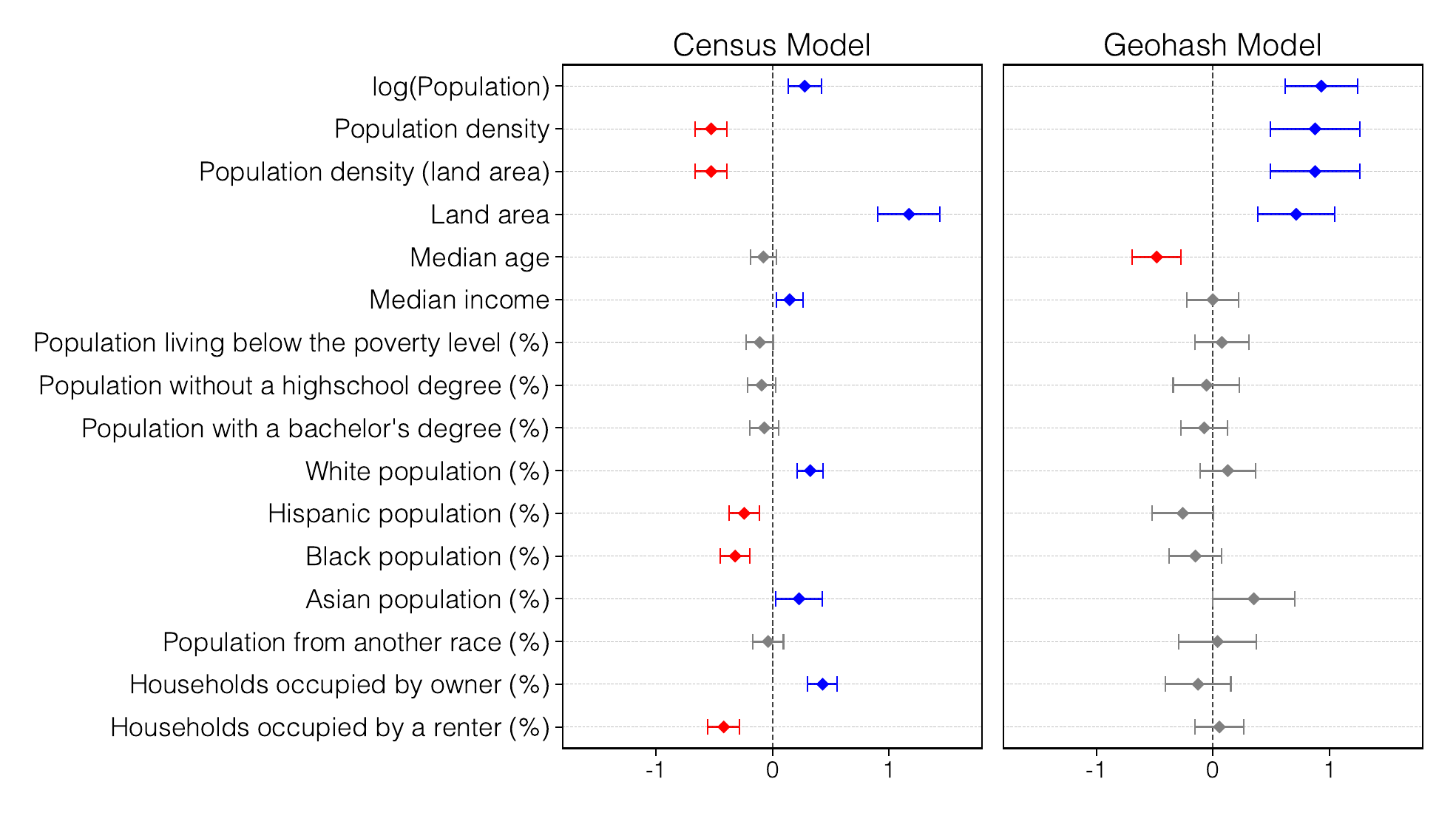}
	\caption
	{Estimated effect of each demographic feature on reporting rate, according to the pooled posterior for each of the networks. Covariates were standardized prior to fitting the model.}
	\label{fig:SI-univariates}
\end{figure}

\Cref{fig:SI-univariates} shows the pooled results for the census tract and the geohash networks. Patterns from \Cref{fig:MAIN-psi} re-emerge on census tracts: reports are heavily skewed along racial lines for census tracts, where a higher percentage of White or Asian residents correlates positively with report probability but a higher percentage of Hispanic or Black residents correlates negatively. Similarly, neighborhoods with higher populations, higher median incomes, and higher proportion of owner-occupied households tend to report events at higher rates than other areas.

One difference between the models is the coefficients for population density (either population normalized by total node area or just by the land area). The census tract model learns negative coefficients for these values, which may be against intuition (more people who can report the same flooded area should increase reporting probability). This negative coefficient shows a limitation of our current Bayesian model: it assumes that the prior probability that each node is flooded $P(A)$ is the same -- even if, e.g., some nodes are larger (so more potential area to be flooded) or in lower lying areas. Thus, if a larger land area (lower population density) is more likely to flood \textit{ conditional on the flooding reports of neighbors}, the model fits this via higher reporting rates. As geohashes have extremely similar land areas, this effect does not appear in the geohash model (and so population density and population data are very similar). Model changes that incorporate such differential priors as a function of node characteristics are a step for future work; we note that all our primary results, as reported in the main text, are consistent between the geohash and census data preprocessing.

\begin{figure}[ht]
	\centering
	\begin{subfigure}{\textwidth}
		\centering
		\includegraphics[width=\textwidth]{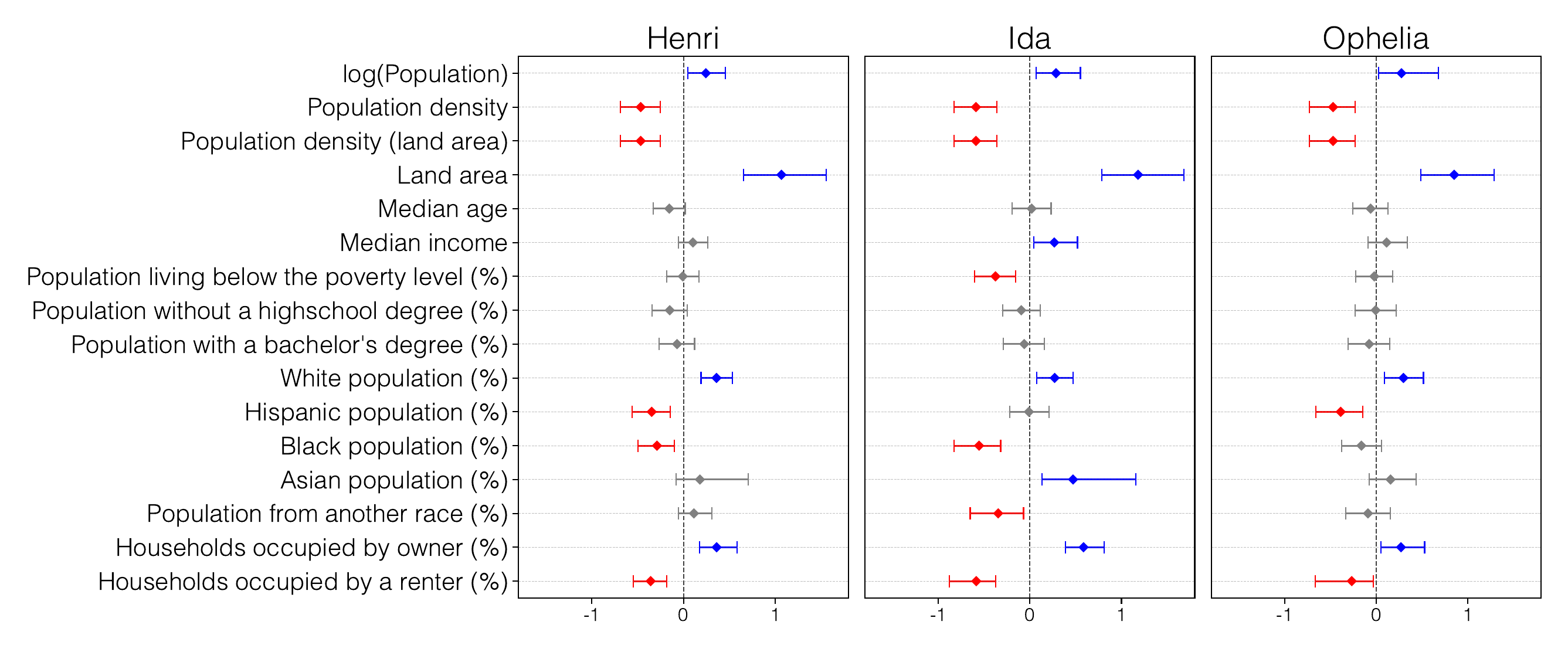}   \caption{Results for the \textbf{census tract} network models}
		\label{fig:SI-univariates-storm-census}
	\end{subfigure}
	\hfill
	\begin{subfigure}{\textwidth}
		\centering
		\includegraphics[width=\textwidth]{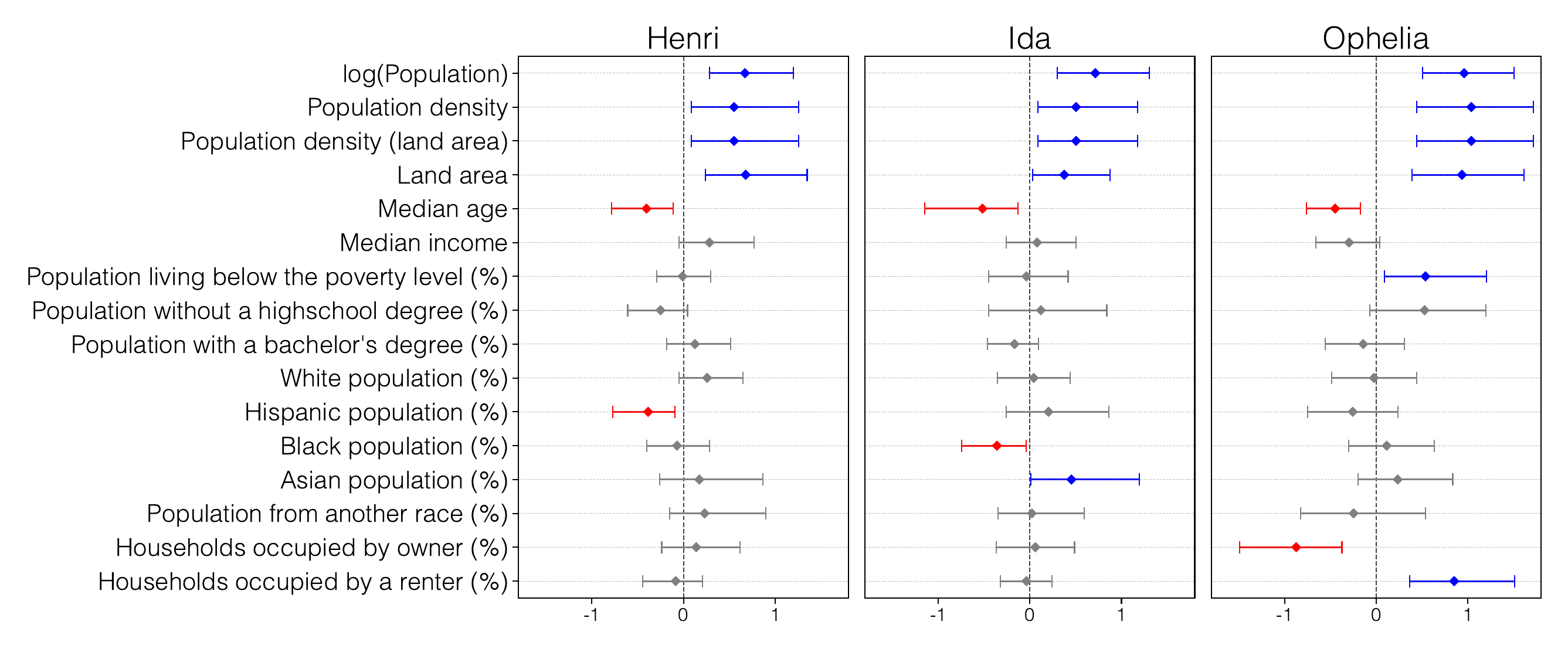}
		\caption{Results for the \textbf{geohash} network models}
		\label{fig:SI-univariates-storm-geohash}
	\end{subfigure}
	\caption
	{Estimated effect of each demographic feature on reporting rate for each storm according to the (a) census tracts and (b) geohash models.}
	\label{fig:SI-univariates-storm}
\end{figure}

Overall, other results directionally agree between the census tract and geohash models. However, most of the geohash correlations are not statistically significant. We attribute this to (1) the lower network size in the case of geohashes, and so fewer data for the associated parameter estimation, and (2) the fact that geohash geometries are not delineated to consider some demographic coherence such as ensuring similar population counts or respecting physical and geographic boundaries---resulting in noisy measurements. These univariate results are consistent across storm---although noisier on the geohash setting, as expected---as shown by \Cref{fig:SI-univariates-storm}.

\clearpage
\FloatBarrier
\section{Further Details on Semi-Synthetic Simulation Experiments}
\label{sec:SI-synthetic}

We now report additional results with our semi-synthetic simulations. Recall that these experiments use the real census tract network and associated demographic features. However, instead of using real report data $\vec{T}$, we simulate ground-truth $\vec{A}$ and report data $\vec{T}$, \textit{assuming} that either the homogeneous reporting or heterogeneous reporting models are correctly specified. In the former case (drawing data according to the homogeneous reporting model), we sample $\theta_0, \theta_1, \alpha$ from their prior distributions, and then use those values to draw $A_i$, $T_i$ for all nodes $i$. In the latter case (drawing data according to heterogeneous reporting model), we sample $\theta_0, \theta_1, \{\alpha_\ell\}$, and then draw $A_i$, $T_i$.

\subsection{Calibration and Identifiability}

\begin{figure}[h]
	\centering
	\includegraphics[width=0.6\textwidth]{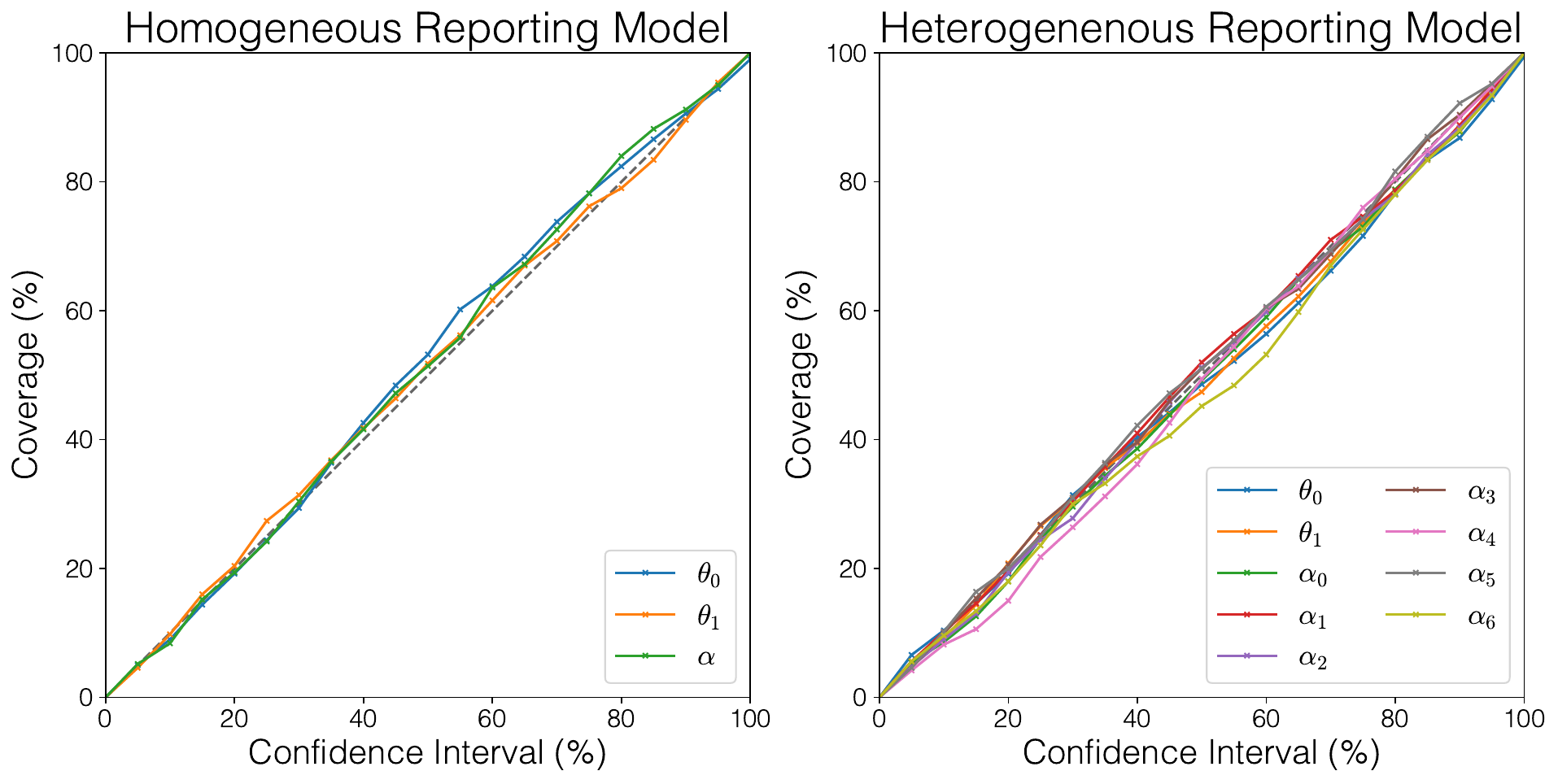}
	\caption{Calibration of synthetic experiments with each of the homogeneous and the heterogeneous reporting models. Point $(x, y)$ corresponds to the $x\%$ confidence interval of the parameter's posterior distribution containing $y \%$ of the ground truth parameters. A perfectly calibrated model lies along the diagonal line.}
	\label{fig:SI-calibration} 
\end{figure}

\begin{figure}[ht]
	\centering
	\begin{subfigure}{.8242\textwidth}
		\centering
		\includegraphics[width=\textwidth]{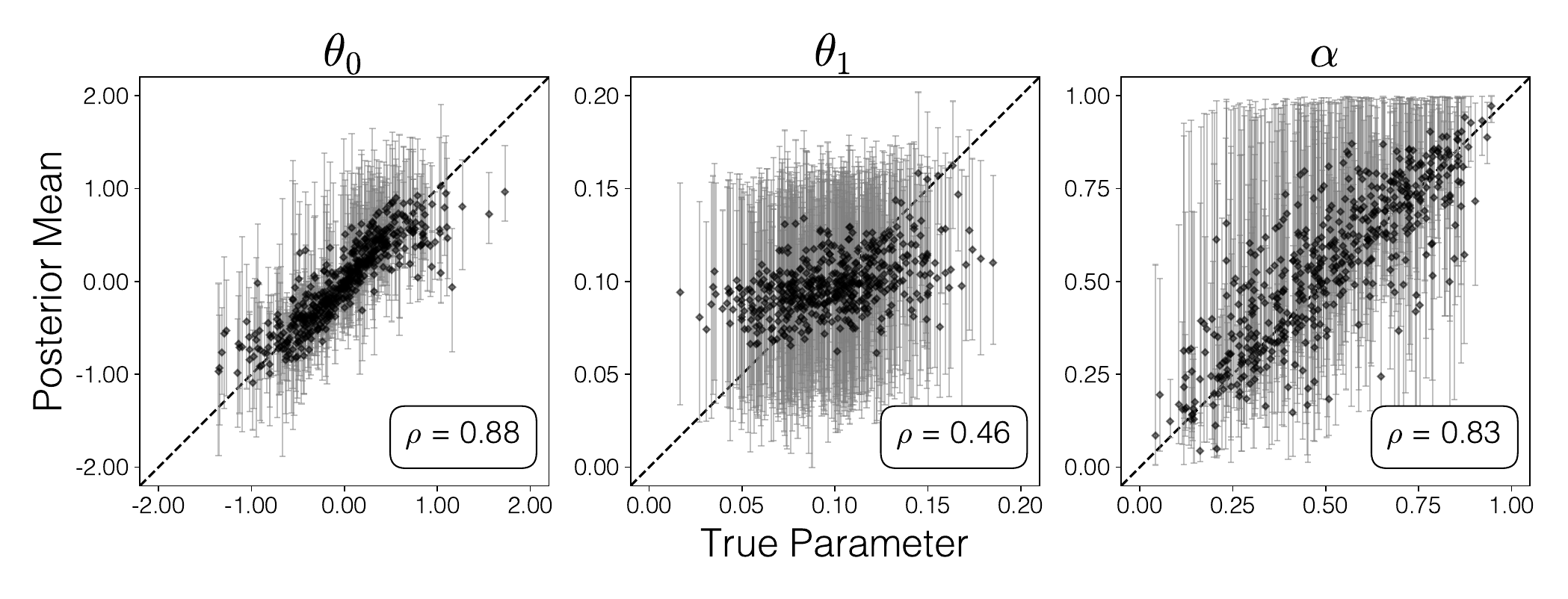}   
		\caption{Results for the \textbf{homogeneous reporting} model}
		\label{fig:SI-identifiability-homogeneous}
	\end{subfigure}
	\hfill
	\begin{subfigure}{.8242\textwidth}
		\centering
		\includegraphics[width=\textwidth]{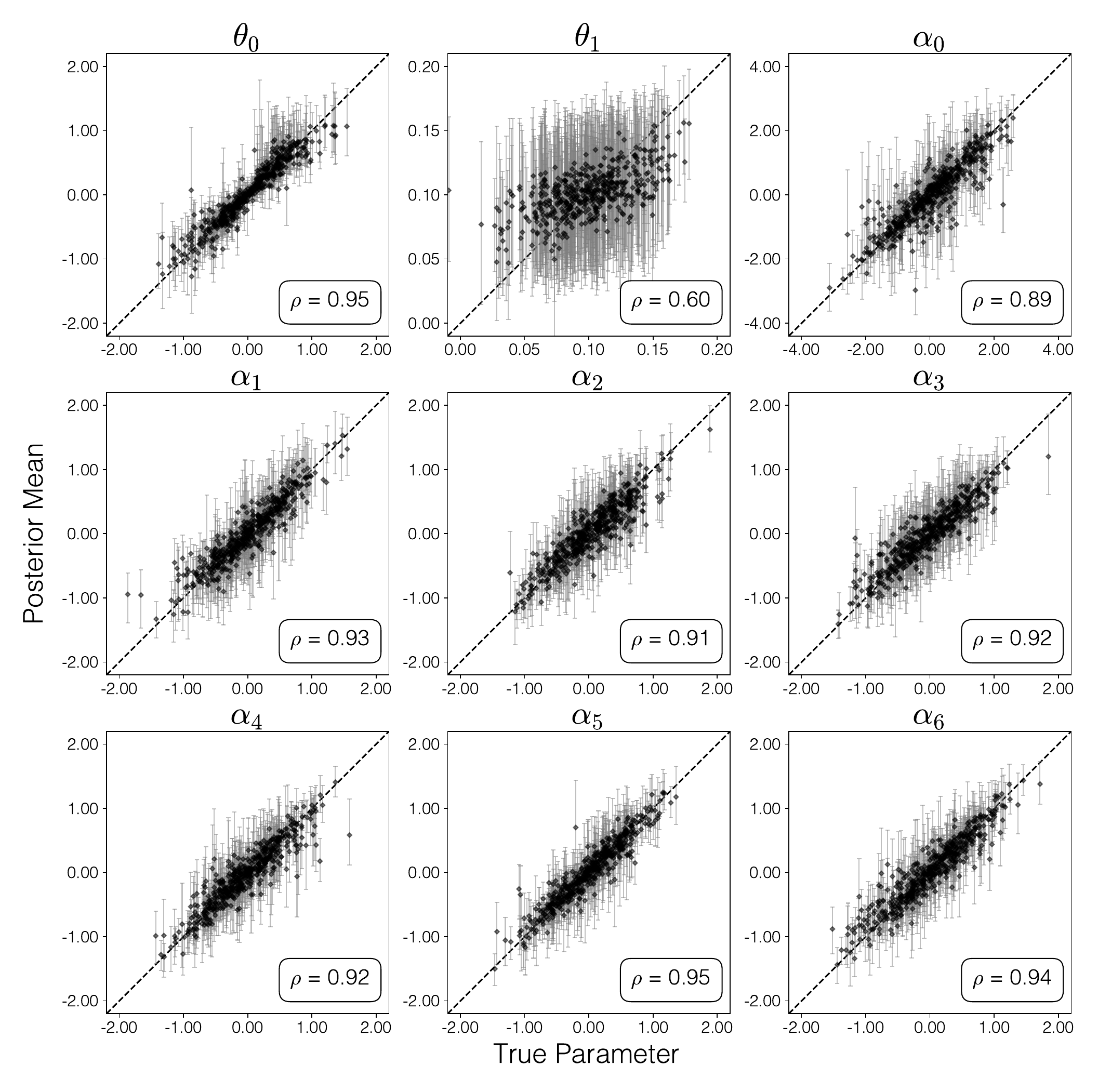}
		\caption{Results for the \textbf{heterogeneous reporting} model}
		\label{fig:SI-identifiability-heterogeneous}
	\end{subfigure}
	\caption
	{Recovery of parameters in simulation in the well-specified (a) homogeneous and (b) heterogeneous reporting models. Vertical axes corresponds to inferred parameter posterior, and horizontal axes to the true parameter value---so that a perfect model would lie along the diagonal. Error bars correspond to $95\%$ confidence intervals of the posteriors.}
	\label{fig:SI-identifiability}
\end{figure}

First, we illustrate the calibration and identifiability findings. \Cref{fig:SI-calibration} shows that both models approach perfect calibration on all latent parameters. We then report parameter identifiability results when the simulated data is drawn from the homogeneous model, and from the heterogeneous model. As the Pearson correlation coefficients $\rho$ illustrate in \Cref{fig:SI-identifiability}, the inferred posterior means are well-correlated with the true parameters. Accurate estimation of the $\alpha_\ell$ is especially important, as these parameters correspond to relative reporting behavior as a function of demographics.

\subsection{Predictive Performance}

We report the AUC and RMSE for our semi-synthetic simulations. We measure the ability of our model to recover \emph{the latent ground-truth} $A_i$ of each node. In all evaluations, nodes that did report during the training period (i.e. $T_i=1$) are excluded as these nodes are perfectly predicted by all models by definition.

\Cref{table:SI-performance_synthetic_homogeneous} shows results for data generated from the homogeneous reporting model i.e. assuming a constant report rate throughout the graph. All three models over-perform random classification, with evidence that our model---the only one which directly accounts for under-reporting---performs better than the two baseline models.

\begin{table*}[hb]
	\centering
	\begin{subtable}[t]{\textwidth}
		\centering
		\begin{tabular}{c|c|c|c|c}
			\textbf{Model} & \textbf{AUC Estimate} & \textbf{AUC 95\% CI} & \textbf{RMSE Estimate} & \textbf{RMSE 95\% CI}\\
			\hline
			Homogeneous Reporting & 0.577 & (0.573, 0.580) & 0.400 & (0.390, 0.410)\\
			GP Baseline & 0.536 & (0.533, 0.539) & 0.494 & (0.482, 0.506)\\
			Spatial Baseline & 0.558 & (0.554, 0.561) & 0.463 & (0.450, 0.477)\\
		\end{tabular}
		\caption{AUC and RMSE point estimates and confidence intervals for each of the three models.}
	\end{subtable}
	\newline
	\newline
	\begin{subtable}[t]{\textwidth}
		\centering
		\begin{tabular}{c|c|c|c|c|c}
			\textbf{Model A}& \textbf{Model B}& \textbf{$\Delta_{\text{AUC}}$} & \textbf{AUC p-value} & \textbf{$\Delta_{\text{RMSE}}$} & \textbf{RMSE p-value}\\
			\hline
			Homogeneous Reporting & GP Baseline & 0.041 & $< 10^{-4}$ & -0.094 & $< 10^{-4}$\\
			& Spatial Baseline & 0.019 & $< 10^{-4}$ & -0.063 & $< 10^{-4}$\\
		\end{tabular}
		\caption{AUC and RMSE changes. The p-values correspond to a two-sided test that Model A and Model B perform differently.}
	\end{subtable}
	\caption{Performance metrics when the data is generated with \textbf{homogeneous under-reporting}. Confidence intervals were obtained through bootstrapping the tracts with 10,000 iterates.}
	\label{table:SI-performance_synthetic_homogeneous}
\end{table*}

When we simulated data according to the heterogeneous reporting model, we additionally compared the performance of the (correctly specified) heterogeneous reporting model to the homogeneous reporting model and the baselines. \Cref{table:SI-performance_synthetic_heterogeneous} shows that accounting for difference in report rates leads to significant improvements on predicting latent ground truth.

\begin{table*}[h]
	\centering
	\begin{subtable}[t]{\textwidth}
		\centering
		\begin{tabular}{c|c|c|c|c}
			\textbf{Model} & \textbf{AUC Estimate} & \textbf{AUC 95\% CI} & \textbf{RMSE Estimate} & \textbf{RMSE 95\% CI}\\
			\hline
			Heterogeneous Reporting & 0.642 & (0.637, 0.647) & 0.378 & (0.369, 0.386)\\
			Homogeneous Reporting & 0.520 & (0.515, 0.525) & 0.464 & (0.450, 0.478)\\
			GP Baseline & 0.501 & (0.497, 0.505) & 0.512 & (0.499, 0.524)\\
			Spatial Baseline & 0.513 & (0.509, 0.518) &  0.487 & (0.472, 0.501))\\
		\end{tabular}
		\caption{AUC and RMSE point estimates and confidence intervals for each of the three models.}
	\end{subtable}
	\newline
	\newline
	\begin{subtable}[t]{\textwidth}
		\centering
		\begin{tabular}{c|c|c|c|c|c}
			\textbf{Model A}& \textbf{Model B}& \textbf{$\Delta_{\text{AUC}}$} & \textbf{AUC p-value} & \textbf{$\Delta_{\text{RMSE}}$} & \textbf{RMSE p-value}\\
			\hline
			Heterogeneous Reporting & Homogeneous Reporting & 0.123 & $< 10^{-4}$ & -0.087 & $< 10^{-4}$\\
			& GP Baseline & 0.142 & $< 10^{-4}$ & -0.134 & $< 10^{-4}$\\
			& Spatial Baseline & 0.129 & $< 10^{-4}$ & -0.109 & $< 10^{-4}$\\
			\hline
			Homogeneous Reporting & GP Baseline & 0.019 & $< 10^{-4}$ & -0.030 & $< 10^{-4}$\\
			& Spatial Baseline & 0.007 & 0.042 & -0.022 & 0.025\\
		\end{tabular}
		\caption{AUC and RMSE changes. The p-values correspond to a two-sided test that Model A and Model B perform differently.}
	\end{subtable}
	\caption{Performance metrics when the data is generated with \textbf{heterogeneous under-reporting}. Confidence intervals were obtained through bootstrapping the tracts with 10,000 iterates. The AUC estimates are repeated from main text.}
	\label{table:SI-performance_synthetic_heterogeneous}
\end{table*}

\clearpage
\FloatBarrier
\section{Further Details on the Equity Allocation Analysis}
\label{sec:SI-equity}

\begin{figure}[h]
	\centering
	\includegraphics[width=\textwidth]{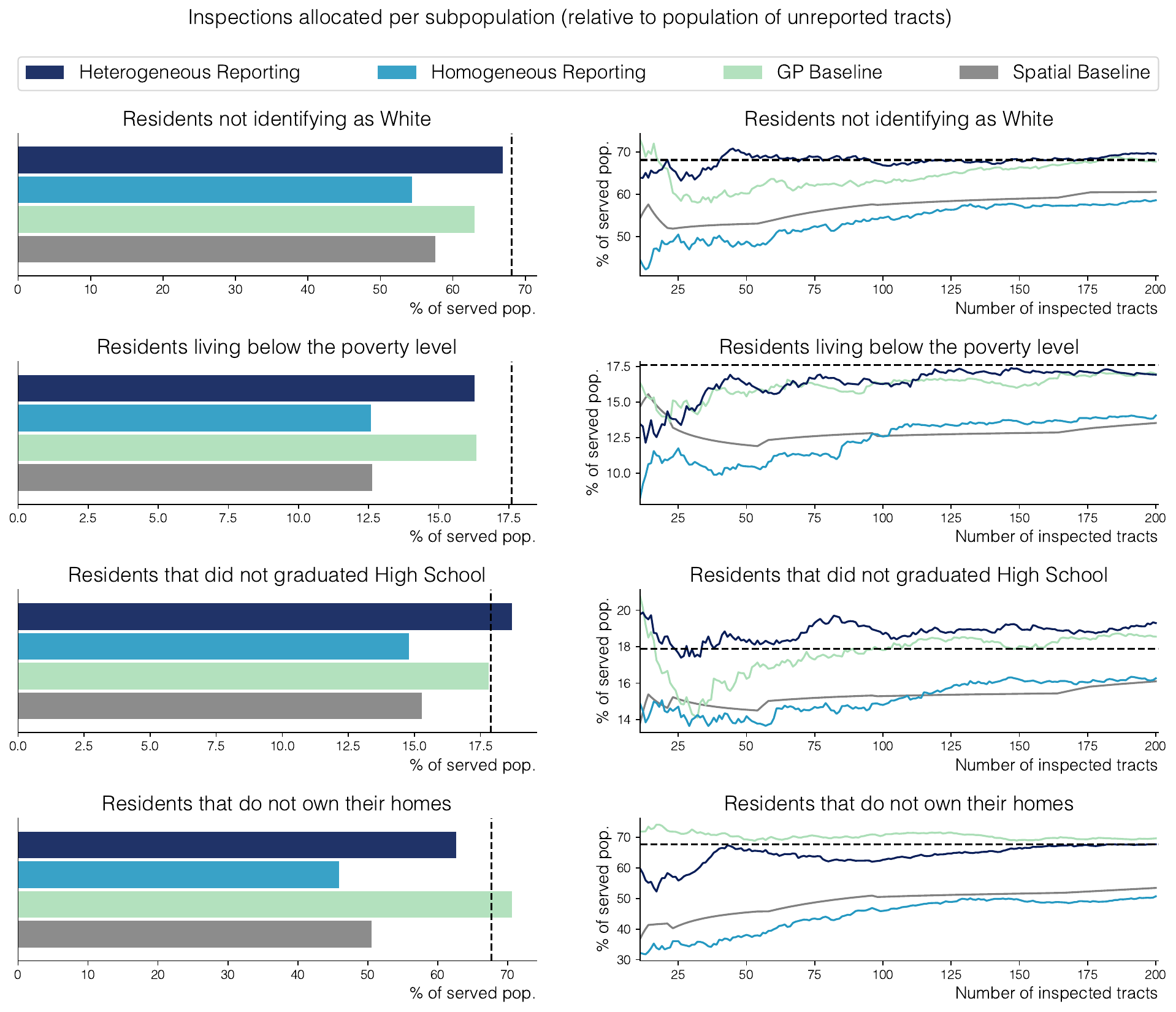}
	\caption{Percentage of non-white, below poverty level, without a high school degree, and household renter populations that are served by inspections of $100$ unreported census tracts---along with a line plot of these values for varying numbers of inspected tracts. Dashed lines correspond to the underlying sub-population percentage living on tracts that did not receive a report.}
	\label{fig:SI-lineplots}
\end{figure}

Qualitatively, the improvements in equity are independent of the number of inspected Census tracts and the socioeconomic or demographic factors considered. We showed in \Cref{fig:MAIN-equity} that the heterogeneous reporting model allocates inspection resources to un-reported census tracts more in line with the population distribution than the homogeneous reporting model along the lines of race and education, when the agency hypothetically would inspect 100 tracts in order of highest posterior probability of a flood. In \Cref{fig:SI-lineplots}, we extend this result to more socioeconomic factors as well as different numbers of inspected census tracts.

Regardless of the number of inspected tracts (after an initial, noisy period), the sub-population rates inspected according to the heterogeneous model and the GP baseline better approach the underlying population base rate than the other models. The GP baseline does so while achieving poor overall performance---as discussed in \Cref{table:MAIN_performance_Ida} (i.e., inspecting tracts close to at random)---whereas the heterogeneous reporting model simultaneously achieves high performance and equity.

Note also that the lines for the spatial baseline model are smoother due to tie-breaking: many tracts have the exact same fraction of their neighbors with a report, and when that happens there is a chance we cannot fit tracts with the same priority in the inspection capacity. We weigh the tracts at the last position equally to fill all the leftover spots.

\clearpage
\FloatBarrier
\section{Further Details on Model Parameters}
\label{sec:SI-model}

\subsection{Model Priors}

\paragraph{Event prevalence $\theta_0$} We assume a centered normal prior of the form:
\[\theta_0\sim \mathcal{N}\left(0, 0.5\right)\]

\begin{figure}[ht!]
	\centering
	\includegraphics[width=0.77\textwidth]{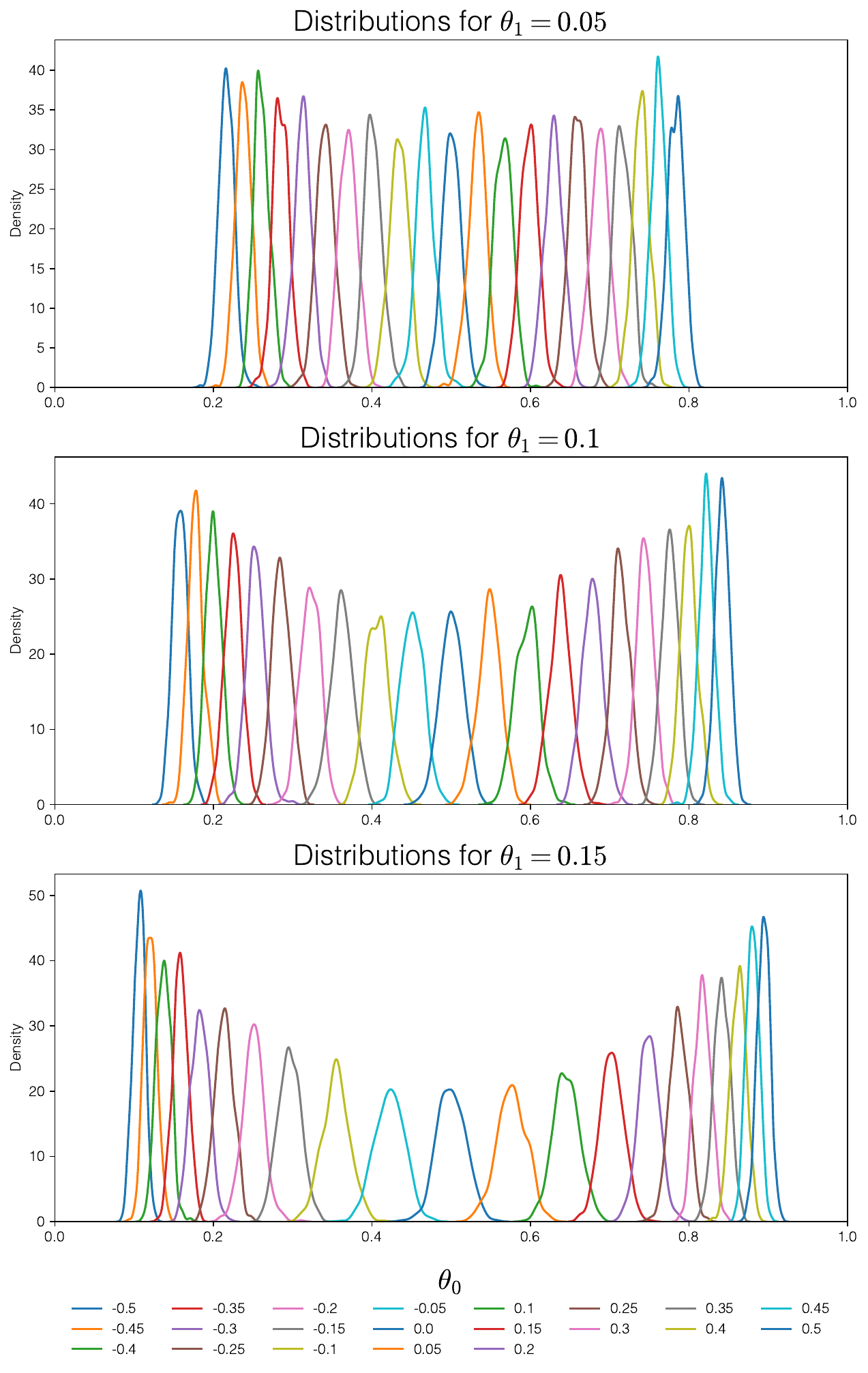}
	\caption{Estimated densities for the fractions of nodes with $A_i=1$ for different choices of $\theta_0$ and $\theta_1$. We generated $500$ samples per pair of parameters.}
	\label{fig:SI-generations-pertheta1}
\end{figure}

\paragraph{Spatial correlation $\theta_1$} We assume a non-negative normal prior of the form:
\[\theta_1\sim \mathcal{N}\left(0.1, 0.03\right)\]
We make two important comments about this prior. First, it forces the spatial correlation to be likely positive; Zero spatial correlation leads to a non-identifiable PU model, and as shown in the data generation experiments, negative spatial correlation---which is not common in the application we leverage---produces a significantly tighter range of latent states $\vec{A}$ regardless of the $\theta_0$ value. Second, although this prior distribution has low standard deviation, the joint distribution of $\theta_0$ and $\theta_1$  generates vectors $\vec{A}$ whose average spans most of the $(-1, 1)$ interval as shown in \Cref{fig:SI-generations-pertheta1}---suitable therefore for real-world data settings.

\paragraph{Regression coefficients $\alpha_\ell$} We assume centered normal prior distributions of the form:
\[\alpha_0 \sim \mathcal N(0, 1.0)\quad\alpha_\ell \sim \mathcal N(0, 0.5)\]

\paragraph{Homogeneous Report Rate $\alpha$} With homogeneous reporting, we impose a loose Beta prior on $\alpha$:
\[\alpha \sim \text{Beta}(1.2, 0.8)\]

We confirm that our prior choices are not \textit{restrictive} -- that they can represent a full range of data. Through \Cref{fig:SI-generations-pertheta1} we show that our spatial correlation $\theta_1$ and location $\theta_0$ parameters can together generate a full range of the fraction of nodes that are positive with the assumed priors.

\subsection{Sampling Hyperparameters}

Each of our chains follows:

\begin{itemize}
	\item The \textbf{number of MCMC iterations} used is 60,000.
	\item The \textbf{number of burn-in iterations} used is 20,000.
	\item The \textbf{thinning fraction of post burn-in MCMC iterations} used is $0.5$.
\end{itemize}

The SVEA is implemented following \citet{moller_efficient_2006}.
\begin{itemize}
	\item The \textbf{number of SVEA iterations} per MCMC iteration is $1$.
	\item The \textbf{number of burn-in samples in the Swendsen-Wang algorithm} used to generate the auxiliary variable is $50$.
	\item The \textbf{initial proposal step size }$\sigma$ on the distribution $\theta_{\text{NEW}}\sim \mathcal{N}(\theta_{\text{OLD}}, \sigma)$ is $0.2$.
	\item The stepsize is adapted \textbf{every $50$ MCMC iterations} if the acceptance rate is not between $0.25$ and $0.60$. The multiplicative factor is $0.15$---either decreasing or increasing the current stepsize, depending on whether we were under-accepting proposals or over-accepting respectively.
	\item The \textbf{parameter initialization} takes the current parameter values for both the $\theta$ parameters and the auxiliary variables (taking current $\vec{A}$ values).
\end{itemize}

The Bayesian logistic regression sampling is implemented with PyMC.
\begin{itemize}
	\item The \textbf{number of burn-in iterations} per MCMC iteration is $50$.
	\item The \textbf{parameter initialization} takes the current parameter values.
\end{itemize}

\clearpage
\FloatBarrier
\section{Further Details on Data}
\label{sec:SI-data}

\begin{figure}[ht!]
	\centering
	\includegraphics[width=\textwidth]{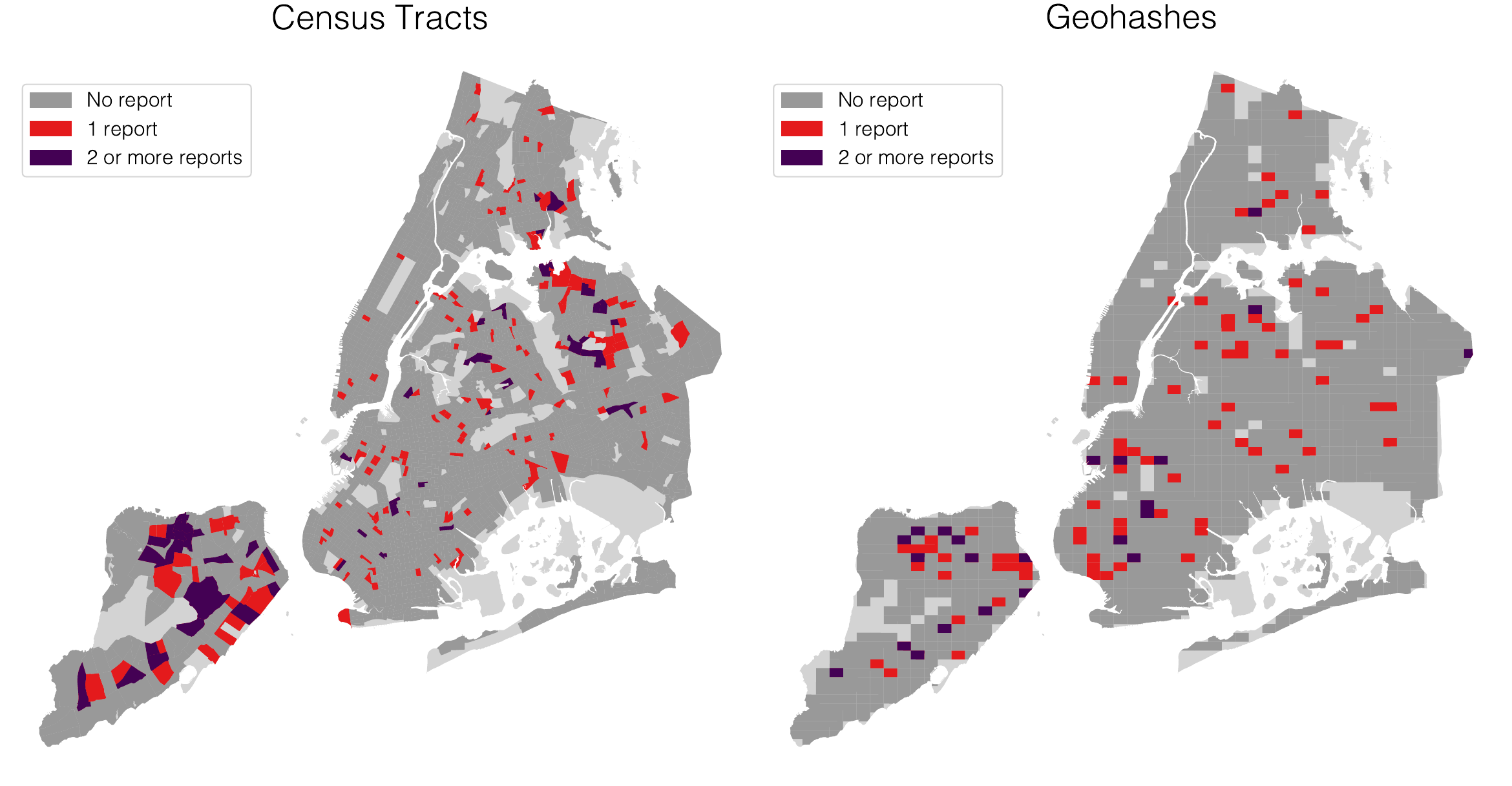}
	\caption
	{Counts of reports during the training period for Hurricane Ida for both networks. Locations in grey received no report. We note that most of the reported tracts have only one or very few reports, validating our use of only the presence of any reports. Because there is a different number of geohashes and census tracts yet we keep a fixed rate of $8\%$ of units reporting in the training period, some census tracts with reports have no corresponding reported geohashes.}
	\label{fig:SI-reports}
\end{figure}

\begin{table}[ht!]
	\centering
	\begin{tabular}{ |l|c|c|}
		\hline
		\textbf{Feature} & \textbf{Average} & \textbf{sd}\\
		\hline
		log(Population) & 8.162 & 0.519\\
		Population density (per $m^2$) & 0.021 & 0.014\\
		Land area ($m^2$) & 283,996 & 396,845\\
		Median age & 37.9 & 5.8 \\
		Median income & \$ 73,909 & \$ 37,072 \\
		Population living below the poverty level (\%) & 16.4 & 12.2 \\
		Population without a high school degree (\%) & 17.4 & 11.4\\
		Population with a bachelor's degree (\%) & 36.9 & 21.3\\
		White population (\%) & 30.3 & 27.3 \\
		Hispanic population (\%) & 26.9 & 21.0 \\
		Black population (\%) & 21.7 & 25.8 \\
		Asian population (\%) & 15.7 & 17.4 \\
		Population from another race (\%)  & 1.7 & 2.7 \\
		Households occupied by the owner (\%) & 37.26 & 26.0 \\
		Households occupied by a renter (\%) & 61.0 & 26.0 \\
		\hline
	\end{tabular}
	\caption{All demographic and socioeconomic features used in the heterogeneous reporting model. Data was obtained from the 2020 Decennial Census. Land area was computed from the geometry. Percentage columns are normalized by the total population (or households) of the Census tract---and tracts with no population are excluded. All features were standardized prior to model fitting. For the full model, we used six features: log(Population), median age, median income, population with a bachelor's degree, white population, and households occupied by the owner.}
	\label{table:demographics}
\end{table}

\begin{figure}[ht!]
	\includegraphics[width=\columnwidth]{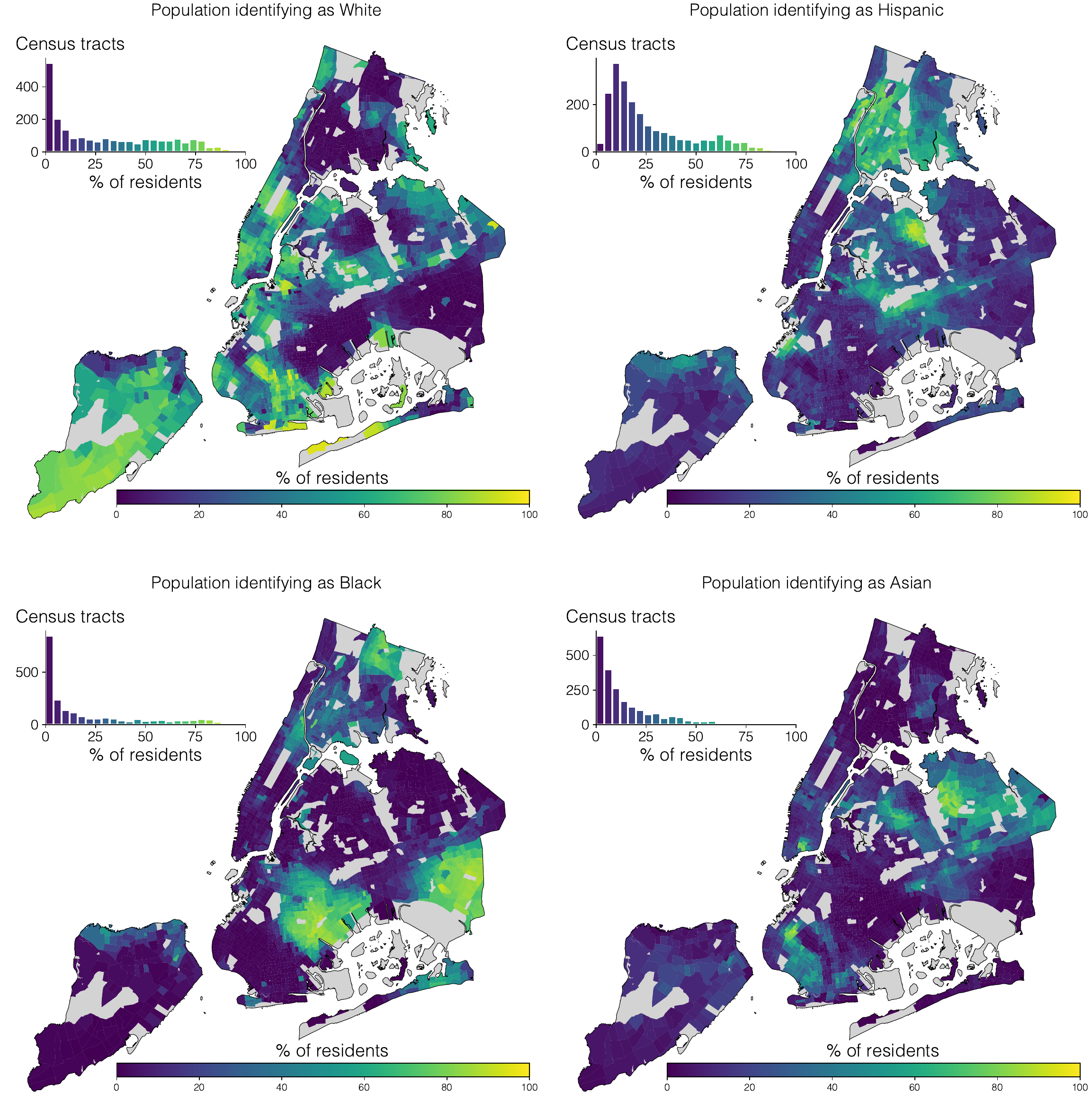}
	\caption
	{Race and Ethnicity in New York City per Census tracts. Each map shows the proportion of tract residents identifying as (a) White (b) Hispanic (c) Black and (d) Asian. As shown, this is a very spatially correlated feature, possibly due to systemic patterns of segregation. As seen in the histograms counting the number of tracts according to their percentage of subpopulation residents, the Hispanic population is more dispersed compared to Black and Asian populations, i.e. few tracts have zero Hispanic residents.}
	\label{fig:SI-race}
\end{figure}

\end{document}